\documentclass[galaxies,article,submit,pdftex,10pt,a4paper]{Definitions/mdpi} 

\firstpage{1} 
\pubvolume{xx}
\issuenum{1}
\articlenumber{5}
\pubyear{2018}
\copyrightyear{2018}
\history{Received: date; Accepted: date; Published: date}

\Title{Progress in Multiwavelength and Multi-Messenger Observations of Blazars
and Theoretical Challenges}

\newcommand{\gtrsim}{\lower3pt\hbox{${\buildrel > \over \sim}$}}
\newcommand{\lesssim}{\lower3pt\hbox{${\buildrel < \over \sim}$}}

\Author{Markus B\"ottcher$^{1}$\orcidA{}}

\AuthorNames{Markus B\"ottcher}

\address{
$^{1}$ \quad Centre for Space Research, North-West University, Potchefstroom, 2520, South Africa;
Markus.Bottcher@nwu.ac.za
}

\corres{Correspondence: Markus.Bottcher@nwu.ac.za; Tel.: +27-18-299-2418}

\abstract{This review provides an overview of recent advances in multi-wavelength 
and multi-messenger observations of blazars, the current status of theoretical
models for blazar emission, and prospects for future facilities. The discussion
of observational results will focus on advances made possible through the {\it Fermi 
Gamma-Ray Space Telescope} and ground-based gamma-ray observatories (H.E.S.S., 
MAGIC, VERITAS) as well as the recent first evidence for a blazar being
a source of IceCube neutrinos. The main focus of this review will be the 
discussion of our current theoretical understanding of blazar multi-wavelength 
and multi-messenger emission, in the spectral, time, and polarization domains. 
Future progress will be expected in particular through the development of the 
first X-ray polarimeter, IXPE, and the installation of the Cherenkov Telescope
Array (CTA), both expected to become operational in the early to mid 2020s.}


\begin{document}

\section{\label{intro}Introduction}

Blazars are the class of jet-dominated, radio-loud active galactic nuclei (AGN) whose relativistic jets
point close to our line of sight. Due to this viewing geometry, all emission from a region moving
with Lorentz factor $\Gamma \equiv (1 - \beta_{\Gamma}^2)^{-1/2}$, where $\beta_{\Gamma}$c is the 
jet speed, along the jet, at an angle $\theta_{\rm obs}$ with respect to our line of sight, will 
be Doppler boosted in frequency by 
a factor $\delta = \left( \Gamma [ 1 - \beta_{\Gamma} \cos\theta_{\rm obs} ] \right)^{-1}$ and
in bolometric luminosity by a factor $\delta^4$ with respect to quantities measured in the co-moving
frame of the emission region. Any time scale of variability in the co-moving frame will be observed
shortened by a factor $\delta^{-1}$. These effects make blazars the brightest $\gamma$-ray sources
in the extragalactic sky \citep[e.g.,][]{3FGL}, exhibiting variability, in extreme cases, on time
scales down to just minutes \citep[e.g.,][]{Aharonian07,Albert07,Aleksic11,Arlen13}. 

The broad-band continuum (radio through $\gamma$-ray) spectral energy distributions (SEDs) of blazars are 
typically dominated by two broad, non-thermal radiation components. The low-frequency component, from radio 
through optical/UV (in some cases, X-rays) is generally agreed to be synchrotron emission from relativistic
electrons in the jet, as evidenced by the measurement of significant, variable linear polarization in 
the radio \citep[e.g.,][]{Lister18} and optical \citep[e.g.,][]{Blinov16} wavebands. For the high-energy 
(X-ray through $\gamma$-ray) SED component, both leptonic (high-energy emission dominated by electrons 
and/or electron-positron pairs) and hadronic (high-energy emission dominated by ultrarelativistic protons) 
models are being considered (for a comparative discussion of both types of models with application 
to a sample of $\gamma$-ray blazas see, e.g., \cite{Boettcher13}, and for a general review of emission
models of relativistic jet sources, see, e.g., \cite{Romero12}). 

The population of blazars consists of flat-spectrum radio quasars (FSRQs) and BL Lac objects, the latter
distinguishing themselves by (nearly) featureless optical spectra with emission-line equivalent widths
$EW < 5 \,$\AA. An alternative classification based on the broad-emission-line luminosity (rather than EW)
was proposed by \cite{Ghisellini11}. The featureless optical continuum spectra of BL Lac objects often 
makes it difficult or impossible to determine their redshift. A more physical distinction between different 
blazar classes might be on the basis of 
the location of their SED peak frequencies. Low-synchrotron peaked blazars (LSP) are defined by having
a synchrotron peak frequency $\nu_{\rm sy} < 10^{14}$~Hz, intermediate-synchrotron peaked blazars (ISP)
have $10^{14} \, {\rm Hz} \le \nu_{\rm sy} < 10^{15}$~Hz, while high-synchrotron peaked blazars (HSPs)
have $\nu_{\rm sy} > 10^{15}$~Hz \citep{Abdo10}. Most HSP and ISP blazars have been classified as 
BL Lac objects based on their optical spectra, while the LSP class contains both 
FSRQs and low-frequency-peaked BL Lac objects. 

In leptonic models, the high-energy emission is produced by Compton scattering of soft (IR -- 
optical -- UV) target photon fields by relativistic electrons. Target photon fields can be the 
co-spatially produced synchrotron radiation \citep[synchrotron self-Compton or SSC; see, 
e.g.,][]{Maraschi92} or external radiation fields, such as those from the accretion disk 
\citep[e.g.,][]{DS93}, the broad-line region \citep[e.g.,][]{Sikora94} a dusty, 
infrared-emitting torus \citep[e.g.,][]{Blazejowski00}, or synchrotron emission from other, 
slower or faster moving regions of the jet, such as a slow sheath surrounding the highly relativistic 
spine in a radially stratified jet \citep[e.g.,][]{Ghisellini05} or another (slower/faster) jet 
component in a decelerating jet flow \citep[e.g.,][]{GK03}. Even though in leptonic models, the 
radiation output is dominated by electrons and/or pairs, it is generally believed that the jets 
also contain non- or mildly-relativistic protons. Due to their much larger mass, they will not 
contribute significantly to the radiative output, but they may still carry a significant (if not
dominant) fraction of the momentum and kinetic power of the jet \citep[see, e.g.,][]{SM00}.

In hadronic models for blazar emission, it is assumed that protons are accelerated to ultra-relativistic
energies so that they can dominate the high-energy emission through proton-synchrotron radiation
\citep[e.g.,][]{Aharonian00,MP01} or through photo-pion production \citep[e.g.,][]{MB92,Mannheim93}, 
with subsequent pion decay leading to the production of ultra-high-energy photons and pairs (and neutrinos!).
These ultra-relativistic secondary electrons/positrons lose their energy quickly due to synchrotron radiation. 
Both these synchrotron photons and the initial $\pi^0$ decay photons have too high energy to escape 
$\gamma\gamma$ absorption in the source, thus initiating synchrotron-supported pair cascade. This typically 
leads to very broad emerging $\gamma$-ray spectra, typically extending into the X-ray regime 
\citep[e.g.,][]{Muecke03}. 

This review will provide, in Section \ref{observations} an overview of recent observational highlights 
on blazars across the electromagnetic spectrum, including aspects of flux and polarization variability 
as well as multi-messenger aspect, especially the recent likely identification of the blazar TXS~0506+056 
as a source of very-high-energy neutrinos detected by IceCube. 
Necessarily, this review will need to focus on recent highlights
for a few selected topics, also biased by the 
author's scientific interests. However, a balanced and fair 
review of relevant works on the selected topics is attempted.

Section \ref{theory} summarizes recent 
developments on the theory side, with a view towards inferences from these recent observational highlights. 
Section \ref{prospects} prospects for future multi-wavelength and multi-messenger observations, especially
towards addressing the following questions, where the author believes that major breakthroughs are possible
through dedicated blazar observations within the next decade:

\begin{itemize}

\item What is the matter composition of blazar jets, and what is the dominant particle population
responsible for the high-energy emission? Answering this question will allow major progress concerning
physics of jet launching and loading, and the mode of acceleration of relativistic particles in jets.

\item What is the structure of magnetic fields in the high-energy emission region and their role in
the acceleration of relativistic particles? The answer to this question will aid in understanding the
physics of jet collimation and stability and provide further clues to the physics of ultra-relativistic
particle acceleration (magnetic reconnection vs. shocks vs. shear layers ...) 

\item Where along the jet is high-energy and very-high-energy $\gamma$-ray emission predominantly produced? 
Different observational results currently point towards different answers (sub-pc vs. 10s of pc from the 
central black hole). The confident localization of the blazar $\gamma$-ray emission region will further 
constrain plausible radiation mechanisms and could possibly hint at beyond-the-standard-model physics
(if evidence suggests that $\gamma\gamma$ absorption in the radiation field of the broad line region is
suppressed below standard expectations). 

\end{itemize}

Throughout the text, physical quantities are parameterized with the notation $Q = 10^x \, Q_x$ in
c.g.s. units.

\section{\label{observations}Recent Observational Highlights}

This section summarizes highlights of recent multi-wavelength and multi-messenger observations of blazars,
focusing on results directly probing the high-energy emission region, which tends to be optically thick to 
radio wavelengths. Radio observations typically probe the larger-scale (pc-scale and larger) structure of
jets and will not be discussed in this review. For recent reviews of radio observations of blazar jets, 
the reader is referred to, e.g., \citep{Gabuzda15,Lister16,Aller17,Jorstad17}.

\subsection{\label{obs_variability}Flux Variability}

Blazars are characterized by their significant variability across the entire electromagnetic spectrum,
on all time scales ranging from years down to minutes. In addition to the very short variability time
scales, a major challenge for our understanding of the optical through $\gamma$-ray emission is the 
fact that the variability patterns in different wavelength bands do not show a consistent behaviour
of correlation (or non-correlation).

\subsubsection{Minute-Scale Variability}

The shortest variability time scales, down to minutes, have been found in very-high-energy $\gamma$-ray 
observations using ground-based Imaging Atmospheric Cherenkov Telescope (IACT) facilities (H.E.S.S., 
MAGIC, VERITAS). They were first identified in HSP blazars, such as PKS 2155-304 \citep{Aharonian07} 
and Mrk 501 \citep{Albert07}, but later also in LSP blazars, namely the prototypical BL Lac object 
BL Lacertae \citep{Arlen13} and the FSRQ PKS 1222+21 \citep{Aleksic11}. Remarkably, sub-hour very-high-energy 
(VHE) $\gamma$-ray variability was also seen by MAGIC in IC 310 \citep{Ahnen17}, which, based on its large-scale
radio structure had been classified as a radio galaxy, in which case relativistic beaming effects
would be inefficient. However, its broadband SED and variability behaviour suggest a blazar-like 
orientation of the inner jet, as is probably also the case for the H.E.S.S.-detected radio galaxy
PKS 0625-35 \citep{Abdalla18}. 

\begin{figure}[H]
\centering
\includegraphics[width=12cm]{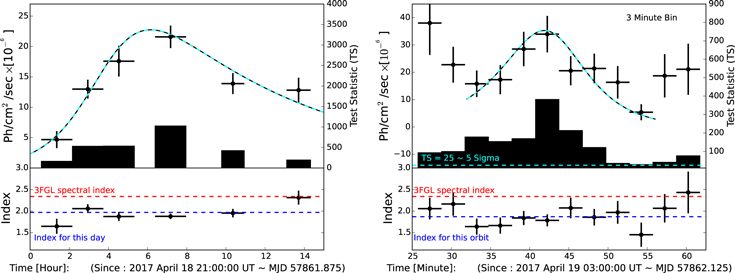}
\caption{{\it Fermi}-LAT GeV $\gamma$-ray light curve of CTA 102 on April 19, 2017. {\it Left}: Orbit-binned
light curve; {\it right}: 3-minute binned light curve \citep[from][]{Shukla18}. Reproduced with permission
from the AAS. }
\label{CTA102minute}
\end{figure}

Rapid variability has also 
been seen at GeV energies with {\it Fermi}-LAT, although the limited photon statistics (because of 
the $\sim 10^5$ times smaller collection area of {\it Fermi}-LAT compared to IACTs) typically limits
the time scales to $\gtrsim$ hours \citep[e.g.,][]{Saito13,Hayashida15}. Note, however, that minute-scale
GeV variability has been detected with {\it Fermi}-LAT during the giang $\gamma$-ray outburst of 3C 279 
in June 2015 \citep{Ackermann16} and, more recently, at $4.7 \, \sigma$ significance during the exceptional 
long-term outburst of the FSRQ CTA 102 in 2016 -- 2017 \citep[][, see Fig. \ref{CTA102minute}]{Shukla18}. 

Due to causality arguments, the observed variability time scale $t_{\rm var}^{\rm obs}$ imposes a 
limit on the size $R$ of the emission region, 

\begin{equation}
R \le c \, t_{\rm var}^{\rm obs} {\delta \over 1 + z} = 1.8 \times 10^{14} \, \left( {t_{\rm var}^{\rm obs} \over 
5 \, {\rm min}} \right) \, \delta_1.
\label{Rvar}
\end{equation}
The observed $\gamma$-ray flux, modulo Doppler boosting, implies a luminosity, which can be translated
into a lower limit on the photon energy density in the emission region, using the size constraint from 
Eq. (\ref{Rvar}). Since there is, so far, no evidence for internal $\gamma\gamma$ absorption on the
co-spatially produced low-energy (IR -- X-ray) radiation field in the 
MeV -- GeV $\gamma$-ray emission of blazars, the emission region must be optically thin to this process. 
This then implies a lower limit on the Doppler factor. $\gamma$-ray photons in the GeV regime interact
primarily with target photons that are observed in the X-ray regime. Assuming an observed X-ray spectrum
with an energy spectral index $\alpha_X$ and an integrated flux $F_{0 - 1}$ in the range $E_0$ to $E_1$
(corresponding to normalized energies $\epsilon_{0,1} \equiv E_{0,1} / [m_e c^2]$), the limit on the 
Doppler factor can be derived as \citep{DG95,Boettcher12}

\begin{equation}
\delta \ge \left( 10^{3 \, \alpha_X} \, {\sigma_T \, d_L^2 \over 3 \, m_e c^4 \, t_{\rm var}^{\rm obs}}
\, F_{0 - 1} \, {1 - \alpha_X \over \epsilon_1^{1 - \alpha_X} - \epsilon_0^{1 - \alpha_X}} \, [1 + z]^{2 \alpha_X} 
\, E_{GeV}^{\alpha_X} \right)^{1 \over 4 + 2 \alpha_X}.
\label{deltamin}
\end{equation}
where $E_{GeV}$ is the maximum photon energy (in units of GeV) out to which there is no evidence for a 
spectral break due to $\gamma\gamma$ absorption (typically of the order of $\sim 10$ -- 100~GeV in the 
case of most Fermi blazars), $d_L$ is the luminosity distance, and $\sigma_T$ is the Thomson cross section. 
In the case of the 5-minute variability
observed in a few blazars, this, in fact, implies minimum Doppler factors of $\delta \gtrsim 50$ 
\citep{Begelman08}. Such values are much higher than the Doppler factors of $\delta \sim 10$ typically 
inferred from superluminal motion speeds observed in radio Very Long Baseline Interferometry (VLBI) 
monitoring observations of blazars 
\citep[e.g.,][]{Hovatta09,Liodakis18} --- a problem sometimes referred to as the {\it Doppler-factor 
crisis} \citep{LL10}. Various suggested model solutions to this problem will be discussed in Section 
\ref{theo_variability}.

\subsubsection{Multiwavelength Correlations}

Another major challenge to our current understanding of the physical processes in blazar $\gamma$-ray emission
regions is the fact that multi-wavelength variability patterns are sometimes correlated, sometimes not, among
different wavelength bands. Even within the same object, the correlated / uncorrelated variability behaviour
changes between different observation periods. Most blazar emission scenarios ascribe the entire IR -- $\gamma$-ray
emission to one single dominant emission region (single-zone models). In this case, one would naturally expect 
all radiating particles to be subject to the same acceleration and cooling mechanisms. As the radiative cooling
time scales of particles are energy-dependent (scaling as $t_{\rm cool} \propto \gamma^{-1}$ for synchrotron
radiation and Thomson scattering), variability patterns at different energies are expected to show time delays, 
but still be correlated. In fact, if an observed time delay between the variability patterns at two frequencies
$E_1 = E_{1, keV}$~keV and $E_2 = E_{2, keV}$~keV is related to different radiative cooling time scales of 
synchrotron-emitting electrons radiating at those energies, this can be used to place a lower limit on the 
magnetic field \citep{Takahashi96}. If electron cooling is dominated by synchrotron and Compton cooling in 
the Thomson regime, a synchrotron time delay $\tau_{1,2} \equiv \tau_{h}$~hr translates into a lower limit 
on the magnetic field of 

\begin{equation}
B \gtrsim 0.9 \, \tau_{h}^{-2/3} \, \delta^{-1/3} \, (1 + k)^{-2/3} \, \left( E_{1, keV}^{-1/2} - E_{2, keV}^{-1/2}
\right)^{2/3} \; {\rm G}
\label{Bcoolinglimit}
\end{equation}
where $k = L_{\rm C} / L_{\rm sy}$ is the Compton dominance parameter of the SED. Such arguments have, in several
cases, constrained the magnetic fields in the emission regions of blazars to be of the order of $B \gtrsim 1$~G
\citep[e.g.,][]{Boettcher03}.

In the case of LSP blazars, leptonic single-zone emission models predict the optical synchrotron emission and GeV 
(Compton) $\gamma$-ray emission to be produced by electrons of approximately the same energy. The optical and 
$\gamma$-ray light curves are therefore expected to be closely correlated with very small time lags. In the case 
of HSP blazars, the same type of correlation is expected to exist between X-rays and VHE $\gamma$-rays. Such a
close correlation is often observed, but there are also many examples in which the correlation is absent. The 
most striking cases are the so-called ``orphan flares'', in which either $\gamma$-ray flares occur with no
visible counterpart in the synchrotron component (\citep[e.g.,][]{Krawczynski04,Blazejowski05}, or synchrotron
(IR -- optical -- X-ray) flares without $\gamma$-ray counterpart.

\begin{figure}[H]
\centering
\includegraphics[width=12cm]{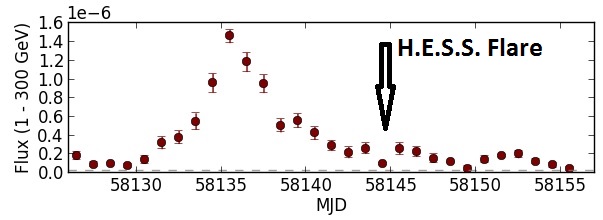}
\caption{{\it Fermi}-LAT GeV $\gamma$-ray light curve of 3C 279 around the flare of January 2018. The 
heavy arrow marks the night of the H.E.S.S. VHE flare detection \citep{JPL2018}}
\label{3C279orphan}
\end{figure}   

A remarkable and unusual case of an orphan flare was detected by H.E.S.S. from the FSRQ 3C 279 on January 27 -- 28, 
2018 \citep{JPL2018}. H.E.S.S. observations had been triggered by a {\it Fermi}-LAT detected GeV $\gamma$-ray flare
of 3C 279 around January 16, 2018. While H.E.S.S. did not detect the source around the time of the {\it Fermi}-LAT 
flare, a significant detecton ($\sim 11 \, \sigma$) resulted with a delay of $\sim 11$~days with respect to the 
GeV flare (see Fig. \ref{3C279orphan}). Also the optical (e.g., from the Steward Observatory Blazar Monitoring
Program\footnote{\tt http://james.as.arizona.edu/~psmith/Fermi/}) and X-ray (e.g., {\it Swift}-XRT\footnote{\tt
https://www.swift.psu.edu/monitoring/}) light curves showed no renewed activity at the time of the H.E.S.S. VHE
flare detection. Such orphan flares appear to strongly argue against simple single-zone 
emission models, and possible alternatives will be discussed in Section \ref{theo_variability}.

\subsubsection{Periodicities?}

AGN activity is often associated with recent galaxy mergers \citep[e.g.][]{DB12,Fu15,Shabala17,Capelo17}. 
If this is true, then one might expect binary supermassive black hole (SMBH) systems, instead of a single SMBH, 
to be present in the centers of at least some AGN. The orbital modulation as well as Lense-Thirring precession
of the dominant accretion disk (and likely also the jet), might then lead to periodic or quasi-periodic modulations 
of the multi-wavelength emissions of blazars. However, the only case in which quasi-periodicity, likely related to
the presence of a binary SMBH system, is clearly established, is the BL Lac object OJ 287, where a dominant SMBH
of $\sim$~a few $\times 10^9 \, M_{\odot}$ appears to be in a $\sim 12$~yr orbit with a smaller SMBH, which intercepts
the primary accretion disk twice per orbit \citep{LV96,Valtonen06a,Valtonen06b}. 

Given the long time lines of many (especially optical and radio) blazar monitoring campaigns and the now 
$\sim 10$~yrs of operations of {\it Fermi}, continuous $\gamma$-ray and well-sampled multi-wavelength 
light curves now exist for a large number of blazars, allowing for efficient searches for periodicities
on yearly time scales. The most promising candidate for such periodicities appears to be the BL Lac object
PG 1553+113, where a period of $\sim 2.2$~years has been identified in {\it Fermi}-LAT and multi-wavelength 
data \citep{Ackermann15}, possibly including secondary ``twin peaks'' symmetrically spaced around the main
peaks \citep{Tavani18}. A systematic search for periodicities in {\it Fermi}-LAT blazar light curves by 
\cite{Covino18}, however, finds no significant evidence for periodicities beyond 95~\% confidence in any 
of the 10 sources studied, including PG~1553+113. Thus, one may conclude that the question concerning
periodicities and the existence of binary SMBHs in blazars (beyond the case of OJ 287) is still controversial.

\subsection{\label{obs_polarization}Polarization --- Variability}

The radio through optical emission from blazars has long been known to be polarized with significant variability
both in the degree of polarization ($\Pi$) and the polarization angle (PA). In this review, we will focus on optical 
polarization measurements, as those might be the best probes of the magnetic field structure in the high-energy 
emission region. The optical polarization in blazars varies from virtually unpolarized sources / states to highly 
polarized states with polarization degrees of $\Pi \lesssim 50$~\% \citep[e.g.,][]{Smith17}. In addition to 
photopolarimetric and spectropolarimetric monitoring of blazars by, e.g., the Steward Observatory blazar monitoring 
program, a systematic study of the polarization variability of a large sample of blazars was performed with the 
RoboPol polarimeter\footnote{\tt http://robopol.org/} on the 1.3m telescope of the Skinakas Observatory 
\citep{Pavlidou14}. 

\begin{figure}[H]
\centering
\includegraphics[width=9cm]{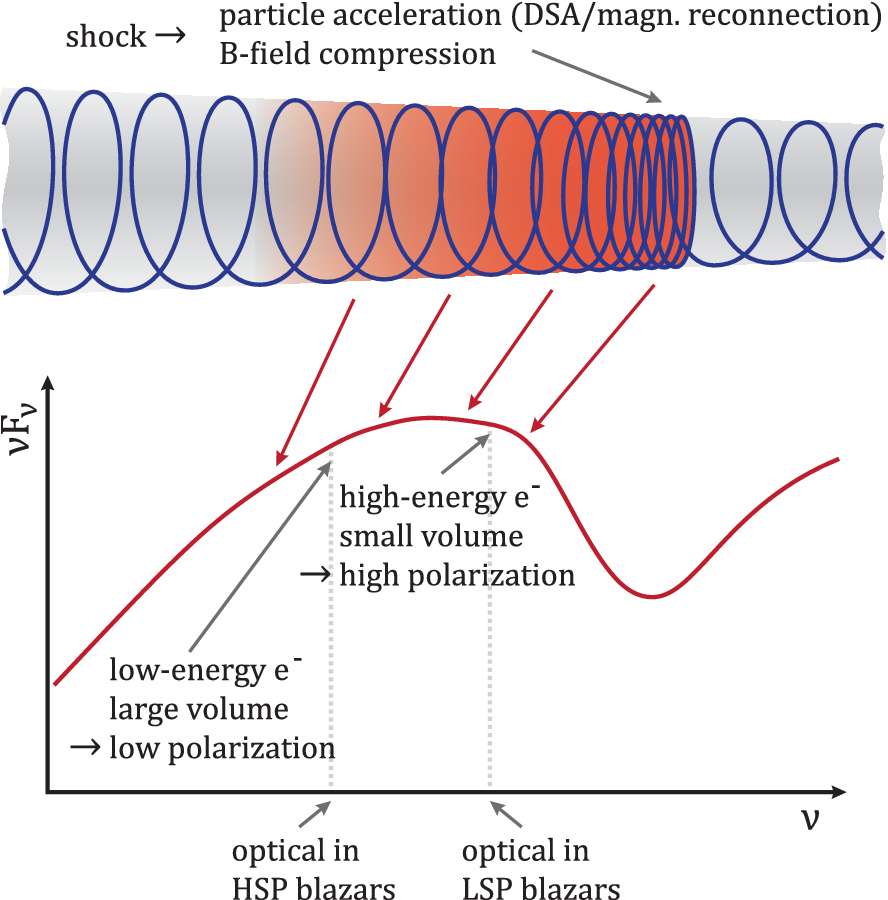}
\caption{Sketch to explain the dependence of the optical PA on the synchrotron peak frequency: In LSP blazars,
the optical spectrum is near the peak of the synchrotron emission, thus reflecting freshly accelerated electrons.
In HSP blazars, the optical range is far below the synchrotron peak frequency, thus reflecting electrons that have
already cooled substantially after the initial acceleration. From \cite{Angelakis16}. Reproduced with kind permission
from Oxford University Press and the Royal Astronomical Society. }
\label{Angelakis2016Sketch}
\end{figure}

The average degree of polarization has been found to be systematically larger for $\gamma$-ray loud blazars 
compared to $\gamma$-ray quiet ones \citep{Angelakis16}, which is likely due to the fact that $\gamma$-ray 
loud blazars appear more strongly Doppler boosted and thus more strongly synchrotron dominated in the optical 
spectrum. The average degree of optical polarization systematically decreases as a function of increasing 
synchrotron peak frequency within the RoboPol sample \citep{Angelakis16}. This may be attributed to the fact that 
in LSP blazars, the optical range is at or near the peak of the synchrotron emission, thus reflecting freshly
accelerated electrons in a presumably very confined region in the jet, while in HSP blazars, optical-synchrotron-emitting
electrons have already cooled substantially and are most likely distributed over a larger portion of the jet
(see Fig. \ref{Angelakis2016Sketch}). 

\begin{figure}[H]
\centering
\includegraphics[width=12cm]{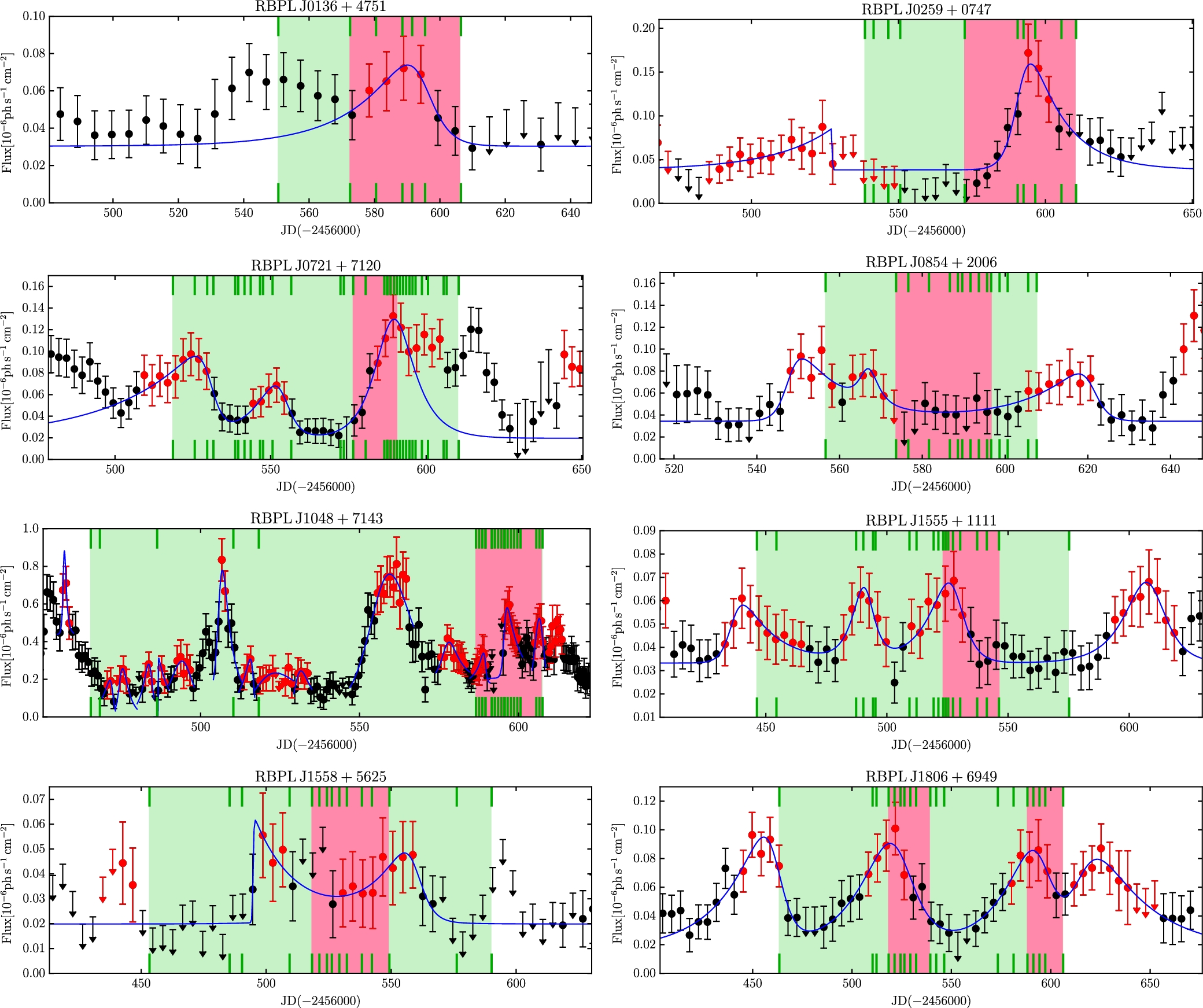}
\caption{{\it Fermi}-LAT $\gamma$-ray light curves of a representative sample of 8 blazars monitored by RoboPol.
The green areas show the period of RoboPol monitoring; red areas indicate periods of PA rotations. 
From \cite{Blinov18}. Reproduced with kind permission from Oxford University Press and the Royal Astronomical 
Society.}
\label{RoboPol}
\end{figure}

While the PAs in blazars typically exhibit erratic small-angle variations, they occasionally undergo systematic 
PA swings exceeding $180^o$, typically over the course of a few days \citep{Marscher08,Abdo10b,Marscher10}. These
PA swings are often associated with $\gamma$-ray and multi-wavelength flares. A central goal of the RoboPol project
was a systematic study of such PA swings in a large sample of blazars 
\citep{Pavlidou14,Blinov15,Blinov16,Blinov16b,Hovatta16,Blinov18}. Key results of this study were that (a) PA rotations
do not occur in all blazars, but whether a blazar shows PA rotations or not, does not appear to depend on its
sub-class (FSRQ / BL Lac object / LSP / HSP) or its average fractional polarization \citep{Blinov16}, and (b) PA
rotations are statistically correlated with $\gamma$-ray flares detected by {\it Fermi}-LAT, while the reverse is
not true, i.e., not all {\it Fermi}-LAT flares in a blazar showing PA rotations are actually associated with such
a rotation \citep[][ see Fig. \ref{RoboPol}]{Blinov18}. Notably, also, there does not appear to be a preferred 
direction of PA rotations, i.e., PA swings can occur in either direction in any given object.

The observations of such PA variability, and especially the PA swings, have spurred a large number of theoretical 
works attempting to explain them. They will be discussed in Section \ref{theo_polarization}.

\subsection{\label{obs_neutrinos}Multi-Messenger Observations --- Neutrinos}

The past few years (2015 -- 2018) marked the birth of multi-messenger astronomy, with the first direct detection
of gravitational waves from a binary-black-hole merger \citep{GW150914}, the first confirmed multi-messenger detection
of gravitational waves from a binary-neutron-star merger and the associated short gamma-ray burst \citep{GW170817},
and the first strong hint for a blazar as the source of astrophysical high-energy neutrinos 
\citep{IceCube18a,IceCube18b}. 

The jets of AGN have long been considered a prime candidate for the sites of acceleration of high-energy cosmic rays
and the production of high-energy neutrinos, as detected by IceCube \citep{Aartsen15,Aartsen16}. Such neutrino emission
is expected in hadronic models for the $\gamma$-ray emission from blazars 
\citep[e.g.,][]{Stecker91,PS92,MB92,Mastichiadis96,Muecke03,Dimitrakoudis12,Murase14}.
However, until 2017, all searches for electromagnetic counterparts of the IceCube astrophysical neutrinos remained
inconclusive \citep[e.g.,][]{Adrian16,Aartsen17}, except for the identification of the blazar PKS B1424-418 as the 
possible source of the PeV IceCube neutrino event HESE-35 (aka ``Big Bird''), which was detected during an extended
multi-wavelength outburst of the blazar in 2012 -- 2013 \citep{Kadler16}.

\begin{figure}[H]
\centering
\includegraphics[width=12cm]{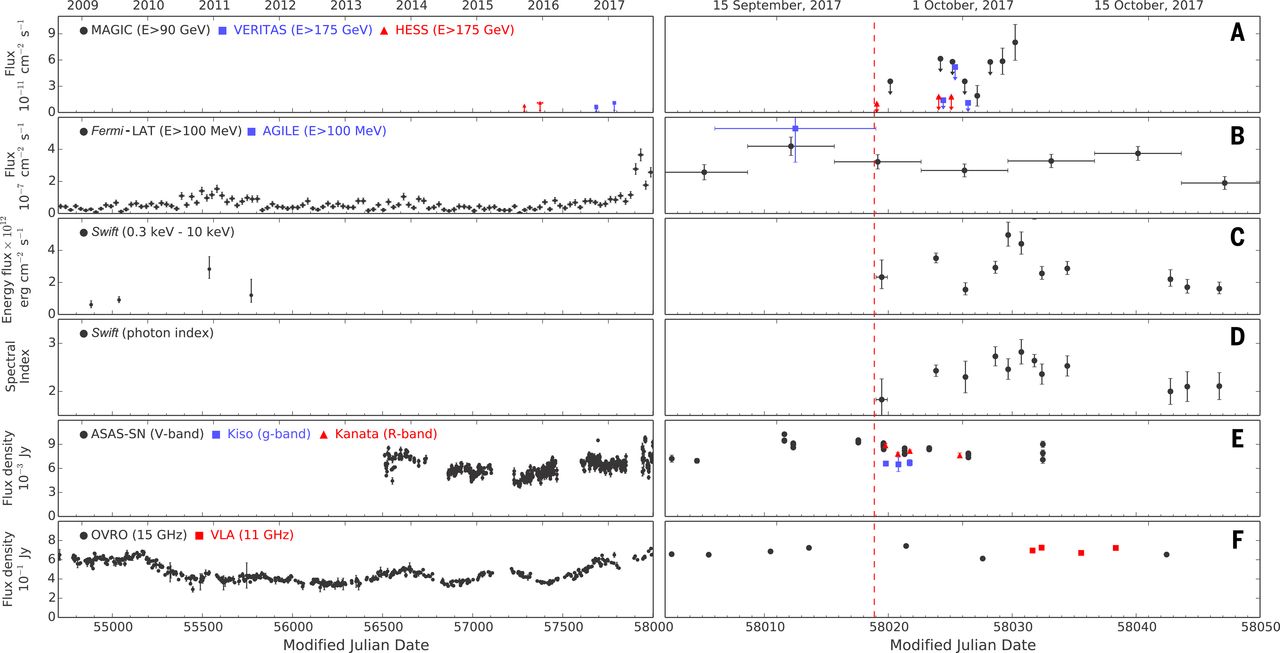}
\caption{Multi-wavelength lightcurves of TXS~0506+056. The vertical red dashed line indicates the time of the 
IceCube-170922A event. Note also the absence of $\gamma$-ray activity during the period 2014 -- 2015 of the 
neutrino flare. From \cite{IceCube18a}. Reproduced with permission by the AAAS. }
\label{TXS0506lightcurve}
\end{figure}   

\begin{figure}[H]
\centering
\includegraphics[width=10cm]{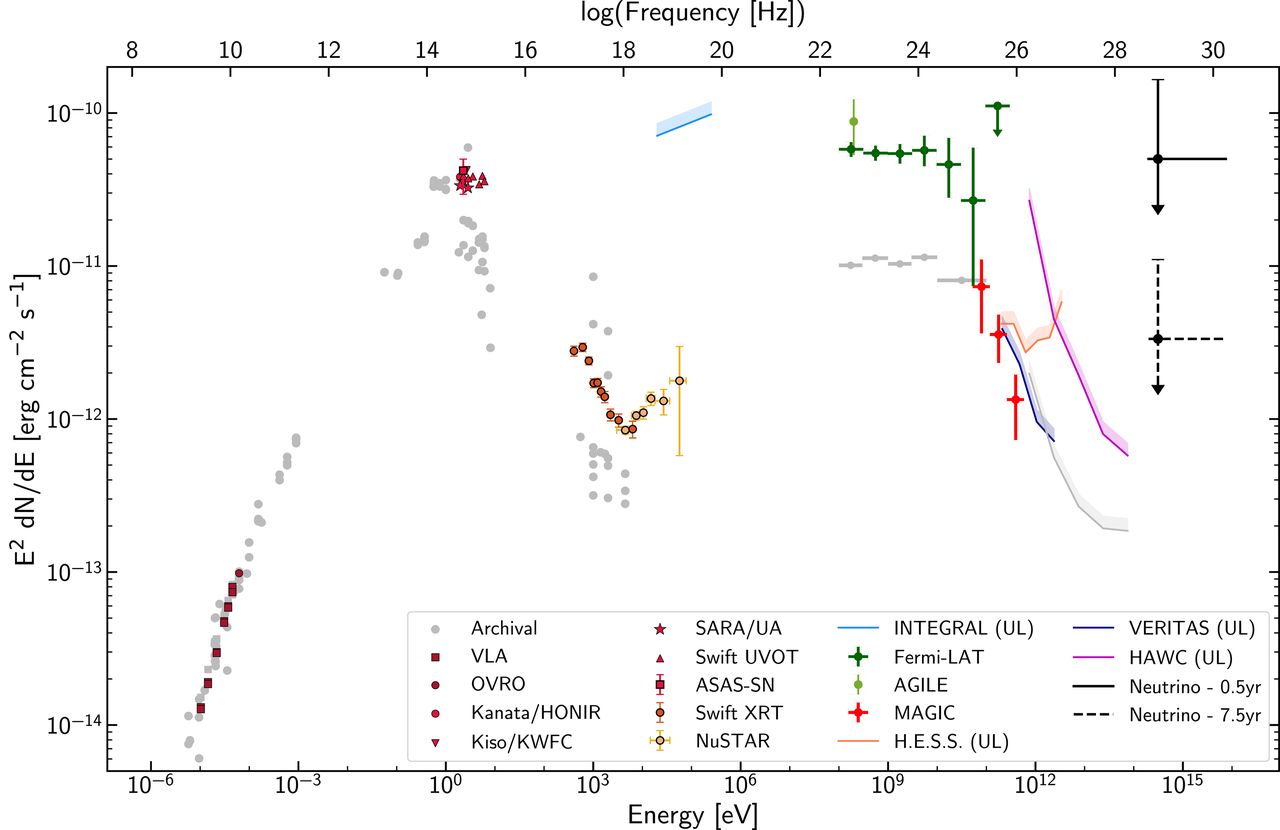}
\caption{Multi-wavelength SED of TXS~0506+056, including neutrino flux upper limits corresponding to the IceCube-170922A
event, assuming one event in 0.5 years (solid black) and one event in 7 years (dashed black). From \cite{IceCube18a}.
Reproduced with permission by the AAAS. }
\label{TXS0506SED}
\end{figure}

\begin{figure}[H]
\centering
\includegraphics[width=10cm]{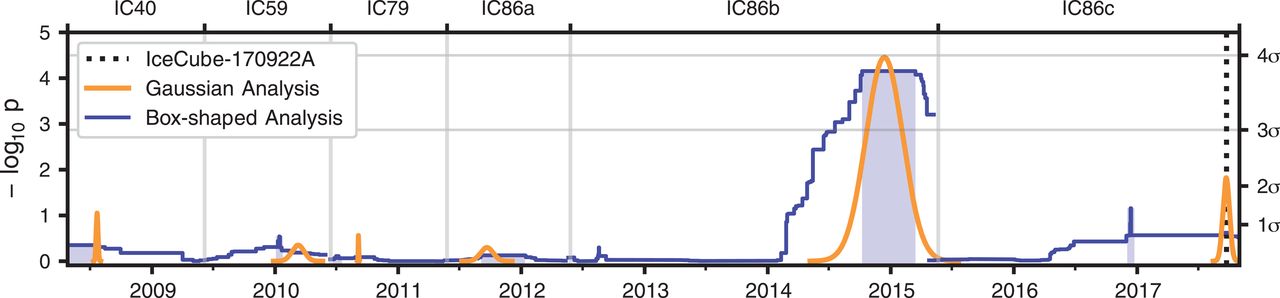}
\caption{Results of a time-dependent search for an IceCube neutrino excess from the direction of TXS~0506+056. 
The orange and blue curves show the results of analyses using a Gaussian and box-shaped time profile for the 
neutrino emission, respectively. The vertical dashed blue line in IC86c indicates the time of the IceCube-170922A
event. From \cite{IceCube18b}. Reproduced with permission by the AAAS. }
\label{TXS0506nuflare}
\end{figure}

This picture changed with the detection of the $\sim 290$~TeV neutrino IceCube-170922A from a direction consistent
with the blazar TXS 0506+056 \citep{IceCube18a} on September 22, 2017. The blazar was in an extended GeV $\gamma$-ray 
flaring state in September -- October 2017 (see Fig. \ref{TXS0506lightcurve}), as detected by {\it Fermi}-LAT, 
and was subsequently also detected in VHE $\gamma$-rays by MAGIC. This single neutrino, however, only had a $\sim 50$~\%
likelihood of actually being of astrophysical origin (due to its moderate energy), thus, by itself, providing only 
marginal evidence for the association. Furthermore, being only one single event, it allowed for the calculation of 
only flux upper limits (see Fig. \ref{TXS0506SED}). In an archival search for additional neutrino events from the
direction of TXS 0506+056, however, the IceCube collaboration found evidence for an excess of $13 \pm 5$ astrophysical
high-energy muon-neutrinos from that location during an extended $\sim 5$~month long period in 2014 -- 2015
\citep{IceCube18b}, henceforth termed the ``neutrino flare'' (see Fig. \ref{TXS0506nuflare}). 
This provided the first strong hint for TXS~0506+056 being a source of high-energy neutrinos, and allowed for the 
first calculation of a measured high-energy neutrino flux from an astrophysical source, corresponding to an 
all-flavour fluence (after correcting for neutrino oscillations) of $4.2^{+2.0}_{-1.4} \times 10^{-3}$~erg~cm$^{-2}$ 
with a spectrum between 32 TeV and 3.6 PeV fitted by a power-law with spectral index $\gamma = 2.1 \pm 0.3$.

\begin{figure}[H]
\centering
\includegraphics[width=10cm]{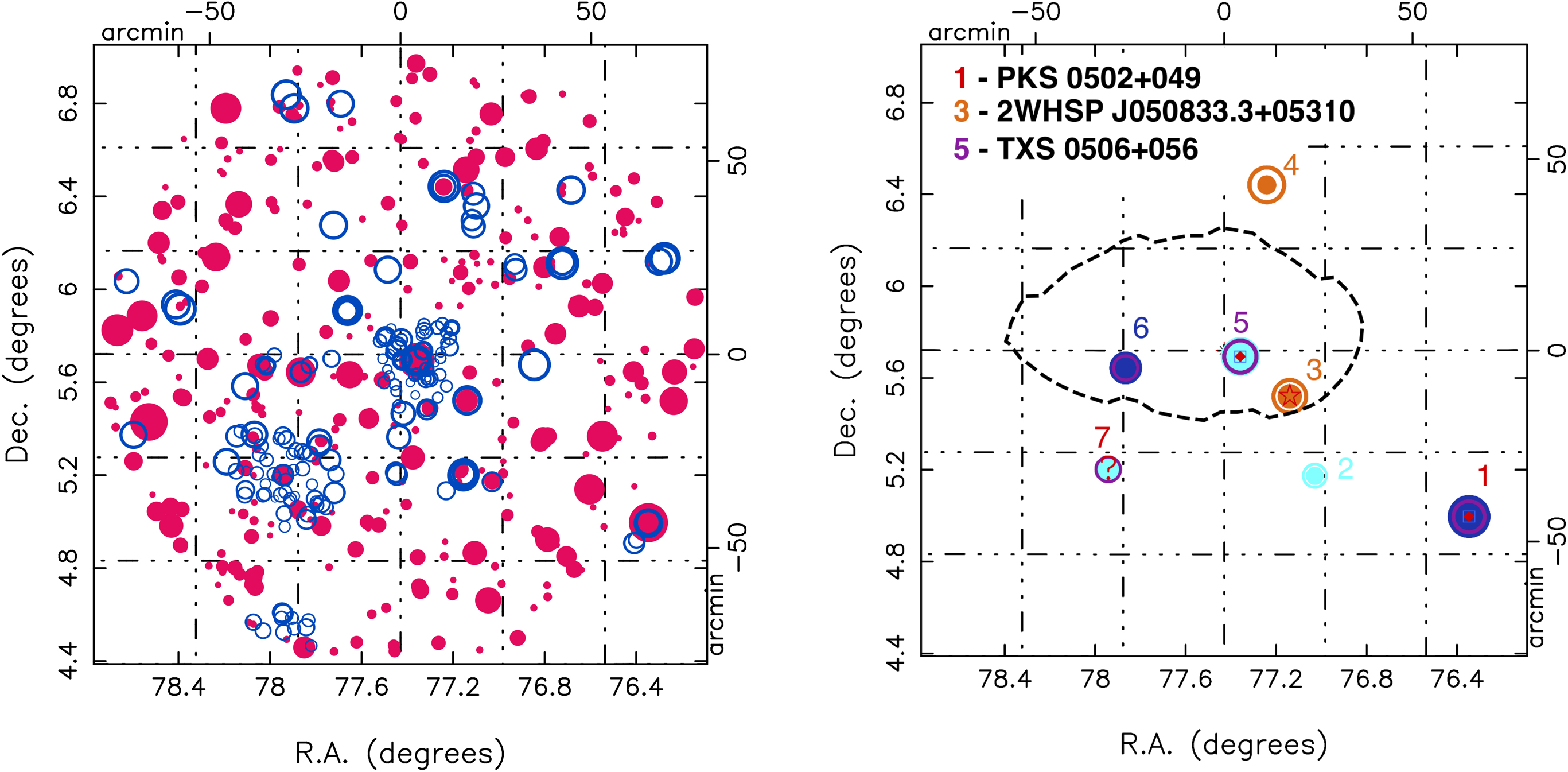}
\caption{{\it Left:} Radio (red) and X-ray (blue) sources within 80 arcmin of the position of the IceCube-170922A 
event. {\it Right}: Known and candidate blazars in the same field. Dark blue circles represent low-frequency peaked
BL Lac candidates, cyan symbols represent intermediate BL Lac objects, and orange symbols indicate high-frequency 
peaked BL Lacs. The dashed line shows the 90~\% error contour of the IceCube-170922 event. From \cite{Padovani18}.
Reproduced with kind permission from Oxford University Press and the Royal Astronomical Society.}
\label{TXS0506region}
\end{figure}

Most notably, the $\gamma$-ray and multi-wavelength flux of TXS 0506+056 during the time of this neutrino flare 
showed no evidence of enhanced activity. Note, however, that \cite{Padovani18} identified a nearby blazar, PKS~0502+049, 
which was in a $\gamma$-ray flaring state for several weeks before and after the 2014 -- 2015 neutrino flare (see Fig. 
\ref{TXS0506region}), but disfavour this source as the potential counterpart of IceCube-170922A as it was not flaring 
during the neutrino flare, but TXS 0506+056 was in a historical hard-spectrum (but low-flux) state at that time. 
The substantial body of theoretical developments spurred by this association will be discussed in Section 
\ref{theo_multimessenger}. 

\section{\label{theory}Theoretical Developments}

This section summarizes some of the most recent developments in the modeling and interpretation of the multi-wavelength
and multi-messenger emission from blazars, specifically addressing the observational highlights described in the previous
section. For a general review of leptonic and hadronic blazar emission models, see, e.g., 
\citep{Boettcher07,Boettcher12,Boettcher13}.

\subsection{\label{theo_variability}Models of Flux Variability}

As elaborated in Section \ref{obs_variability}, observed blazar variability patterns pose at least two major challenges 
to currently existing blazar radiation models: (a) the rapid, minute-scale ($\gamma$-ray) variability, and (b) the 
inconsistent cross-correlation patterns, with emission in different wavelength bands being sometimes correlated, 
sometimes not, including occurrences of orphan flares. This section will provide a brief overview over various 
blazar variability models currently ``on the market'', discussing how they may have the potential to address the 
issues mentioned in the preivous section.

\subsubsection{Causes of variability}

Variability of blazar emission can, in principle, be caused in a variety of ways, by which different models may be
classified:

\begin{enumerate}

\item{{\bf Shock-in-jet models}: In these models (also termed ``internal shock models'') inhomogeneities in the jet 
flow produce mildly relativistic shocks travelling through the (relativistically moving) jet plasma, leading to the 
acceleration of particles, most plausibly through Diffusive Shock Acceleration \cite[DSA; see, 
e.g.,][]{Drury83,BE87,JE91,SB12,Baring17}. The shock-in-jet model for blazars was first suggested in the seminal
work by Marscher \& Gear \citep{MG85}, and subsequently refined in a large number of works, mostly in the framework 
of leptonic emission scenarios 
\citep[e.g.,][]{Spada01,Sokolov04,Sokolov05,Graff08,BD10,JB11,Chen11,Chen12,Joshi14,Zhang14,Zhang15,Zhang16a}.

\begin{figure}[H]
\centering
\includegraphics[width=7.5cm]{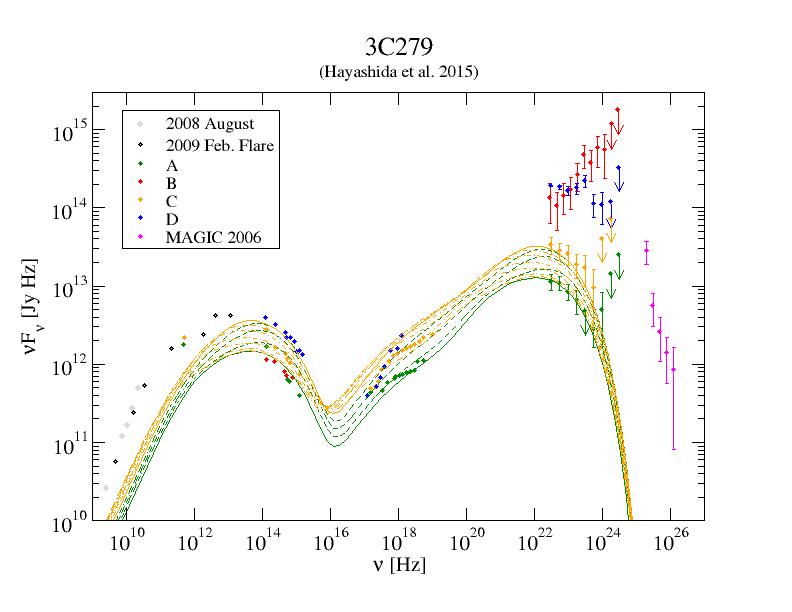} \quad
\includegraphics[width=7.5cm]{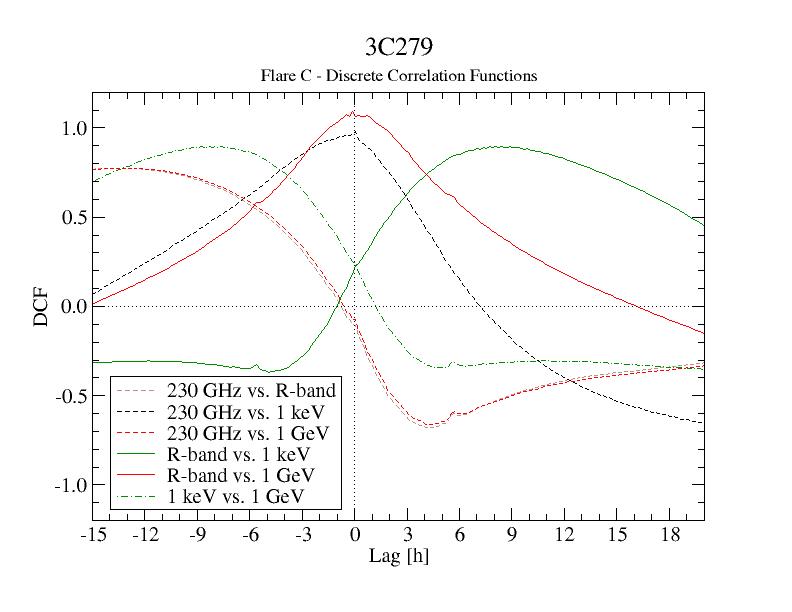}
\caption{Shock-in-jet simulation of a flare of the FSRQ 3C279. Data are from \citep{Hayashida15}. {\it Left}:
Snap-shot SEDs resulting from the time-dependent simulaton; {\it right}: Cross-correlations between various 
frequency bands. Radio and X-ray variations are expected to lag behind optical and GeV $\gamma$-ray variations
by $\sim 10$~hours. From B\"ottcher \& Baring (2019, in preparation). }
\label{3C279_DCF}
\end{figure}   

In shock-in-jet models, particle acceleration occurs in a single region around the shock, equally affecting all 
radiating particles (leptonic or hadronic). Thus, these models generally predict correlated multi-wavelength 
variability with inter-band time lags reflecting energy-dependent electron (or proton) cooling (and/or acceleration)
time scales \citep[e.g.,][]{BD10}. In the case of a leptonic emission scenario, for FSRQs, the optical and 
GeV $\gamma$-ray emissions are produced by electrons of similar energies and therefore are expected to be 
correlated with close to zero time lag. Radio and X-rays are produced by lower-energy electrons, expected to
exhibit a delayed response compared to the optical and $\gamma$-ray emissions. Fig. \ref{3C279_DCF} shows an 
example resulting from a shock-in-jet simulation representing a multiwavelength flare of the FSRQ 3C 279 
(flare C from \citep{Hayashida15}), where the radio and X-ray emissions are expected to lag behind the 
optical and VHE $\gamma$-ray emissions by $\sim 10$~hours. In the case of HSP blazars, the X-ray and $\gamma$-ray 
emissions are expected to be closely correlated, with GeV $\gamma$-ray and optical emissions lagging behind
the X-ray and VHE variations. 

\begin{figure}[H]
\centering
\includegraphics[width=10cm]{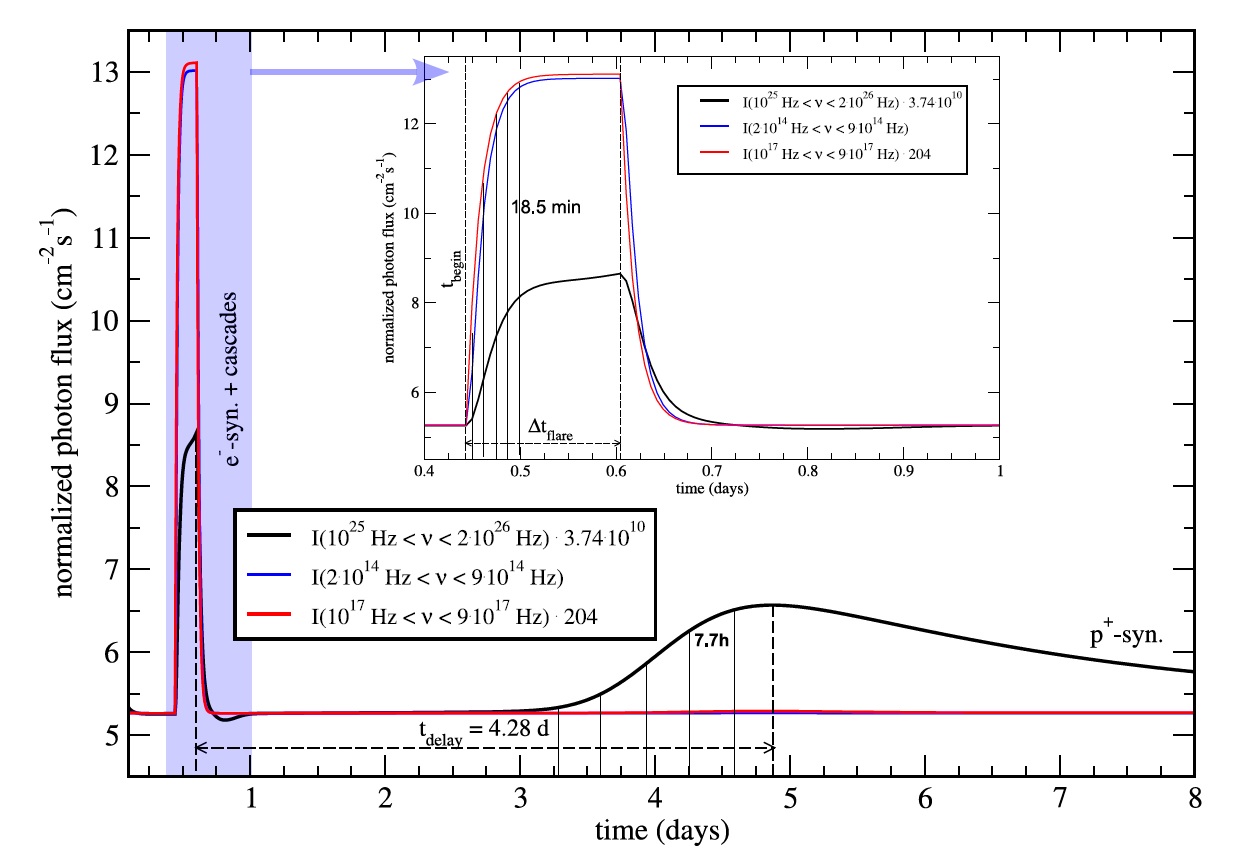}
\caption{Multi-band (optical = blue, X-ray = red, $\gamma$-ray = black) light curves from a lepto-hadronic model
with gradual, stochastic acceleration of particles. The acceleration time scale for protons is substantially longer
than for electrons, leading to a delayed $\gamma$-ray orphan flare due to proton-synchrotron emission. 
From \cite{WS15}. Reproduced with permission from ESO. }
\label{WSorphan}
\end{figure}   

Uncorrelated variability amongst the two SED components could be achieved, in 
the framework of shock-in-jet models, with the assumption of the emission from the dominant shock being strongly 
synchrotron- or high-energy (Compton in leptonic models) dominated, enhancing the quiescent emission from the 
larger-scale jet only in a narrow frequency band \citep[e.g.,][]{KT06,Potter18}.
Alternatively, in a hadronic shock-in-jet scenario, vastly different acceleration time scales of electrons and
protons might give the impression of uncorrelated variability due to the long time delay between the radiation 
components produced by electrons and protons \citep[e.g.,][see Fig. \ref{WSorphan}]{WS15}. 

Shock-in-jet models naturally assume that the shock affects the entire cross section of the jet, with a radius 
of typically $R_{\perp} \sim 10^{15}$ -- $10^{16}$~cm. As discussed in Section \ref{obs_variability}, this 
constrains the variability time scale to $t_{\rm var} \gtrsim R_{\perp} \, (1 + z) / (\delta \, c) \sim 9
\, R_{\perp, 16} \, (1 + z) / \delta_1$~hours, which is very difficult to reconcile with minute-scale variability,
unless a very small jet cross section and/or a very large Doppler factor are assumed. 

It is well known that the formation of strong shocks and efficient particle acceleration at shocks is suppressed 
in the presence of a dominant magnetic field \citep[e.g.,][]{Sironi15b}. The low magnetizations 
($u_B / u_e \sim 0.1$ -- $10^{-3}$) typically inferred from broadband SED modeling of blazars 
\citep[see, e.g.,][]{Ghisellini10,Boettcher13} therefore seem to support the hypothesis of shocks being the 
dominant particle acceleration sites in blazars. 
}

\bigskip

\item{{\bf Turbulence / Magnetic Reconnection:} The relativistic flows of AGN jets are likely to develop 
turbulence, which may trigger magnetic reconnection. This has been studied in a large number of works in 
recent years \citep[e.g.,][]{Kagan15,Sironi15,Nalewajko15,Werner16,Guo16}. It has been shown that magnetic 
reconnection produces hard power-law spectra of relativistic electrons, $n_e (\gamma) \propto \gamma^{-p}$, 
including spectral indices approaching a value of 1 \citep[e.g.,][]{Guo14}, which, however, is also achievable
with oblique relativistic, but still subluminal shocks \citep{SB12}.

\begin{figure}[H]
\centering
\vbox{\includegraphics[width=10cm]{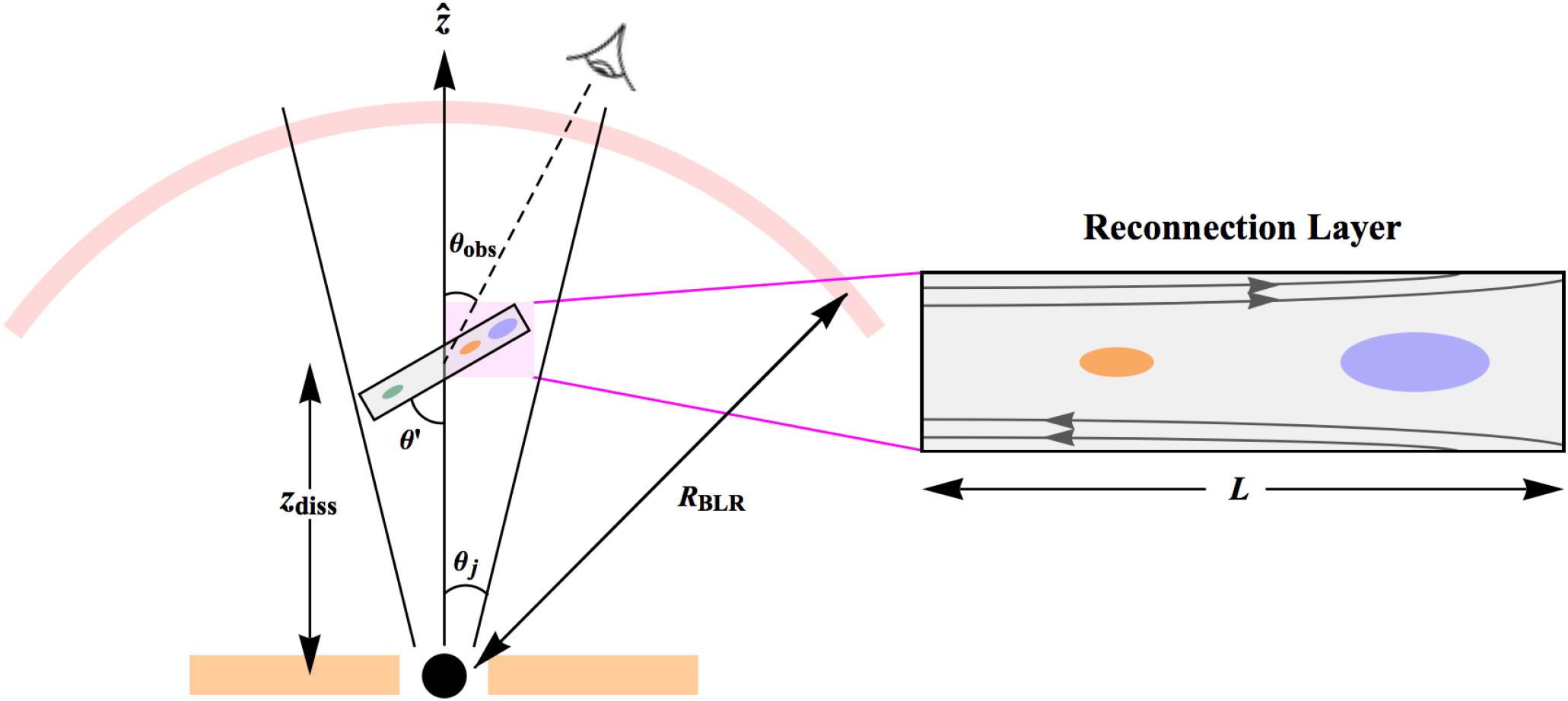} }
\bigskip
\vbox{ \includegraphics[width=10cm]{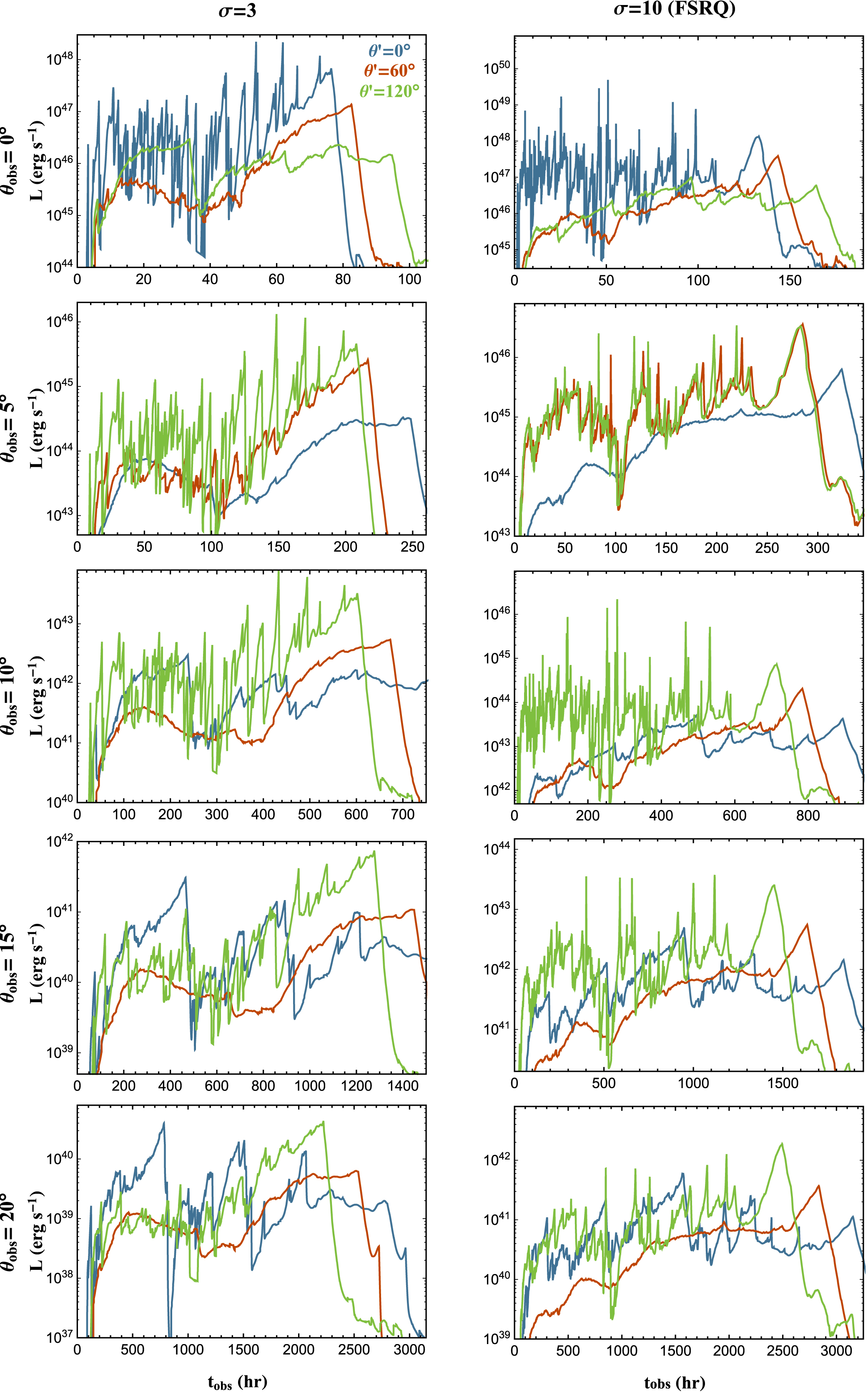} }
\caption{{\it Top:} Sketch of a jet-in-a-jet scenario where particle acceleration results from pasmoid-dominated
magnetic reconnection. {\it Bottom:} Simulated 0.1 -- 300~GeV $\gamma$-ray light curves at different viewing angles 
(top to bottom: $\theta_{\rm obs} = 0^o$, ..., $20^o$) for two different plasma magnetization values of 
$\sigma = 3$ (left) and $\sigma = 10$ (right). Different colours indicate different degrees of alignment of the 
plasmoids with the jet axis. The light curves illustrate the rapid, large-amplitude variability that can be produced 
in such a scenario. From \cite{Christie19}. Reproduced with kind permission from Oxford University Press and the 
Royal Astronomical Society.}
\label{minijet}
\end{figure}

In view of the minute-scale variability problem discussed in Section \ref{obs_variability}, magnetic-reconnection 
models are particularly appealing as they provide the possibility to produce small-scale ultrarelativistic flows 
within the reconnection region, the so-called ``jets-in-a-jet'' scenario \citep{Giannios09,Giannios10,Giannios13}. 
The resulting ultrarelativistic bulk motion of small plasmoids provides additional Doppler boosting, resulting in
very short, bright flares. With modest magnetization ($\sigma \sim$~a few), this model is capable of producing the
observed fast (minute-scale) variability 
\citep[][see Fig. \ref{minijet}]{GT08,Giannios09,Giannios13,Petropoulou16,Christie19} 
without requiring ultrarelativistic ($\Gamma \gg 10$) bulk motions of the entire jet material. 
}

\bigskip

\item{{\bf External Sources of Variability:} Another class of models attributes variability to interactions of
the jet (or the high-energy emission region within the jet) with matter external to the jet, either by direct 
collisions or by means of radiative interactions. 

\begin{figure}[H]
\centering
\includegraphics[width=10cm]{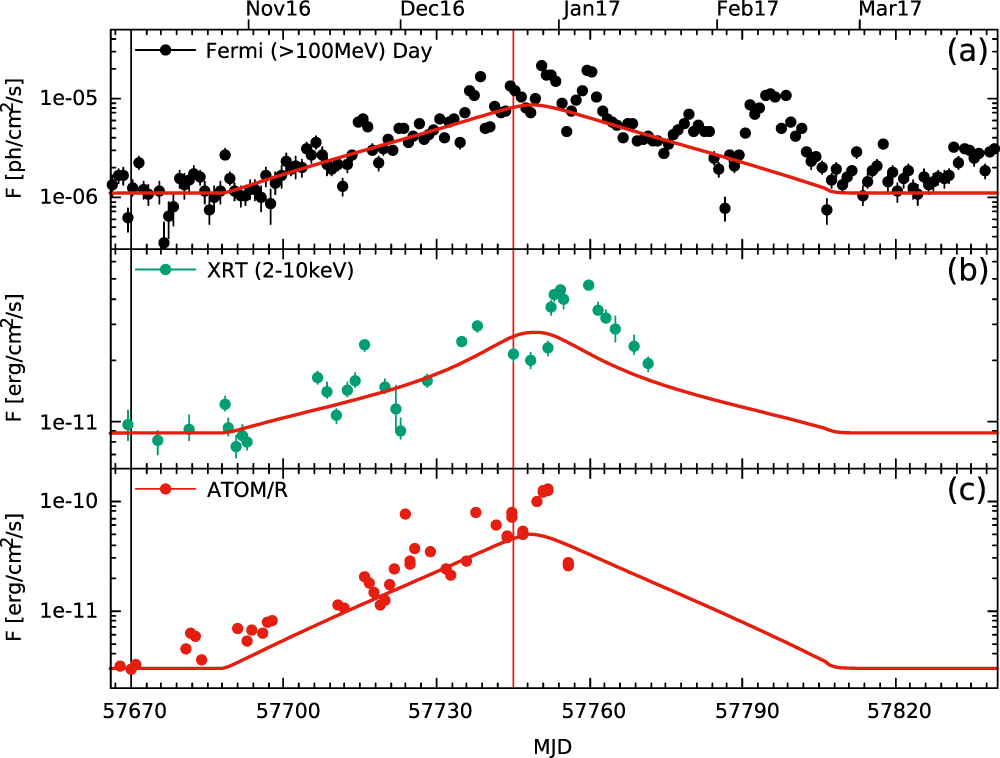}
\caption{Fits to the $\sim 4$~months-long multi-wavelength outburst of CTA 102 in 2016 -- 2017 in $\gamma$-rays
(a), X-rays (b) and optical (c) using a model of cloud ablation by the blazar jet. 
From \cite{Zacharias17}. Reproduced with permission from the AAS. }
\label{CTA102Zacharias}
\end{figure}   

Examples of the former models include jet -- star/cloud collision models 
\citep[e.g.,][]{Araudo10,Barkov10,Barkov12,Araudo13,Khangulyan13}, where the jet interacts with and (at 
least partially) disrupts a star or a gas cloud, leading to the formation of a strong shock with subsequent 
particle acceleration. The natural duration of flares in such a scenario is expected to be of the order of 
the crossing time of the cloud or star through the jet, typically of the order of days to weeks or months.
For example, a model of cloud ablation by a blazar jet has recently been proposed by \cite{Zacharias17}
to model the $\sim 4$~months long giant multi-wavelength outburst of the FSRQ CTA 102 in 2016 -- 2017,
see Fig. \ref{CTA102Zacharias}. Note that \cite{Barkov12} argue that a jet-star interaction may also
produce rapid, minute-scale variability due to the acceleration of fragments of the stellar envelope to
ultra-relativistic bulk speeds.

\begin{figure}[H]
\centering
\includegraphics[width=9cm]{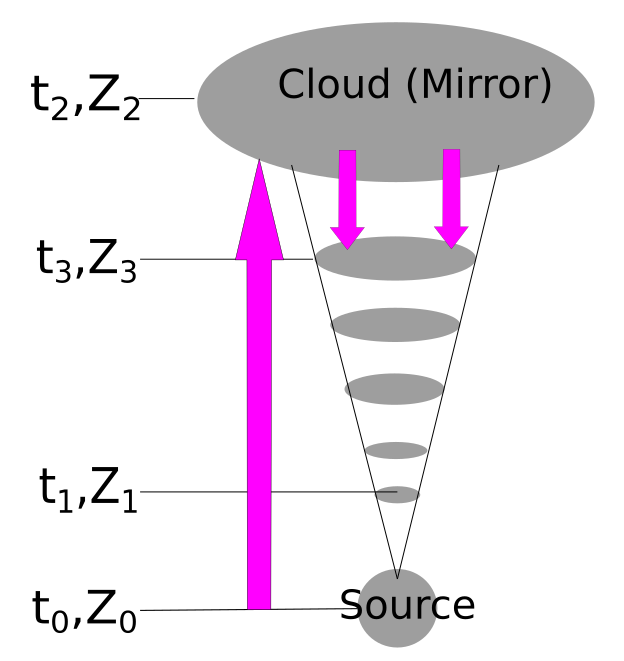}
\caption{Sketch of a synchrotron mirror model. From Oberholzer \& B\"ottcher (2019, in preparation).}
\label{symirrorsketch}
\end{figure}

Variability models based on radiative interactions of the jet with external medium include, in particular,
synchrotron mirror models, in which the synchrotron emission produced within the jet is reflected off an
external obstacle (the ``mirror'', which could, e.g., be a cloud of the Broad Line Region or a stationary
feature within the jet) \citep[e.g.,][see Fig. \ref{symirrorsketch}]{Boettcher05,Tavani15}. It 
thereby appears as an intense target photon field either for Compton scattering (leptonic models -- see Fig. 
\ref{Tavani_lc}) or p$\gamma$ pion producton (hadronic models) for a short time around the passage of the 
high-energy emission region by the mirror. A hadronic synchrotron mirror model has been proposed by
\cite{Boettcher05} to explain the orphan TeV flare of 1ES 1959+650. The reflected synchrotron X-ray 
emission in this case would be an inefficient target for Compton scattering due to Klein-Nishina suppression,
leaving p$\gamma$ interactions off relativistic protons as the dominant signature of the mirror process.

\begin{figure}[H]
\centering
\includegraphics[width=9cm]{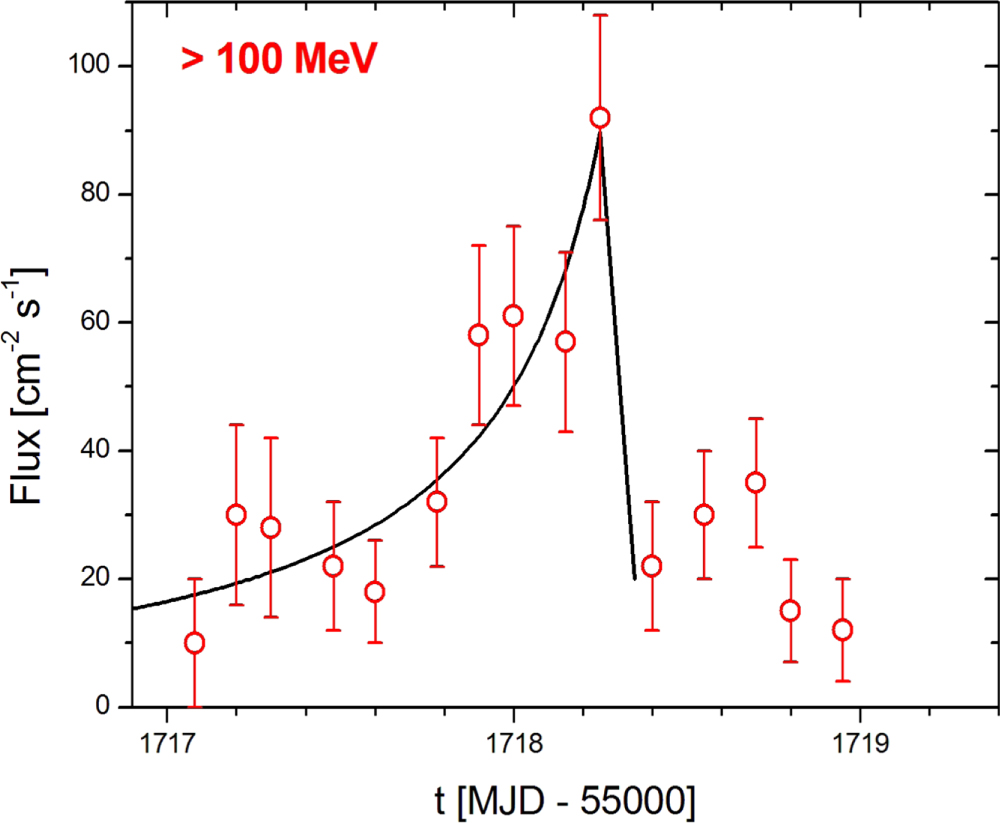}
\caption{Simulated $\gamma$-ray light curve in a leptonic synchrotron mirror model. 
From \cite{Tavani15}. Reproduced with permission from the AAS. }
\label{Tavani_lc}
\end{figure}

Into the same category falls the ``Ring of Fire'' model by \cite{MacDonald15,MacDonald17}, in which the 
high-energy emission region passes a stationary external source of seed photons for Compton scattering
and which has also been proposed as a model for ``orphan flares'' \citep{MacDonald17}. 
}

\bigskip

\item{{\bf Geometric Models:} Variability models based on bending or helical jets invoke a change of the 
viewing geometry as the dominant source of variability, due to a change in the Doppler factor 
\citep[e.g.,][]{VR99,Ostorero04,Larionov13,Larionov16,Raiteri17}. 
Bending or helical jet structures are often observed in radio VLBI monitoring \citep[e.g.,][]{Lister13}. 
Models based on Doppler-factor variability typically predict correlated, almost acromatic variability,
except for a small shift in frequency by the changing Doppler factor, unless one assumes that different 
portions of the electromagnetic spectrum are not produced co-spatially and may be affected by different
Doppler-factor variations \citep[e.g.,][]{Raiteri17}. Such models have also had some success in reproducing
optical polarization variability, as discussed further in the next sub-section. 

Another class of models that may be categorized as geometric invokes particle acceleration triggered
by the kink instability in jets \citep[e.g.,][]{Zhang17,Nalewajko17}. In these models, variability is
caused by particle acceleration due to magnetic energy dissipation in the development of the instability,
which is expected to be accompanied by significant changes in the polarization signatures, as discussed
further below. 
}

\end{enumerate}

\subsubsection{Numerical approaches}

\begin{figure}[H]
\centering
\includegraphics[width=12cm]{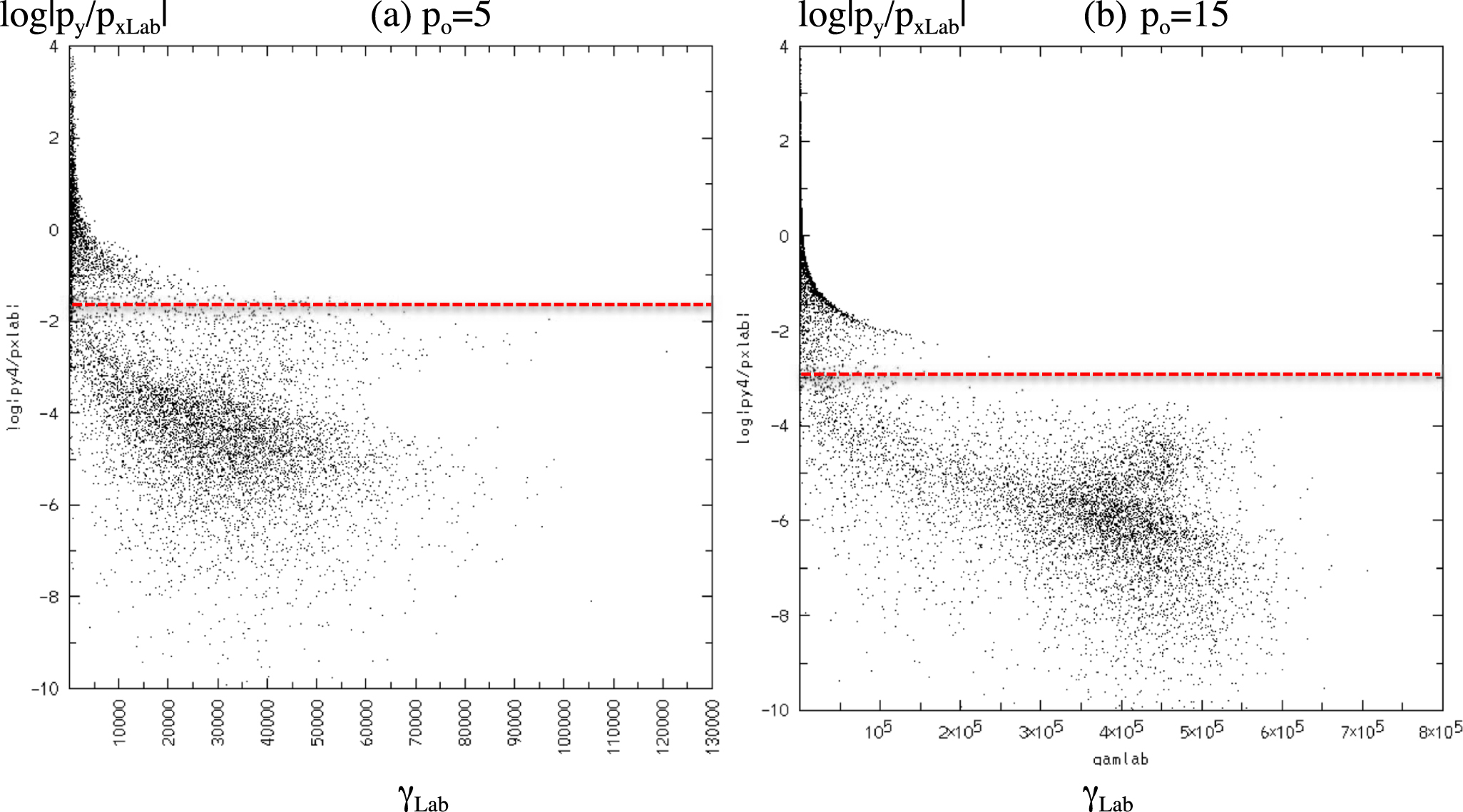}
\caption{Anisotropic particle acceleration in relativistic shear layers from particle-in-cell simulations.
Plotted is the log of the angle of the particles' motion with respect to the jet axis, vs. particle energy, 
indicating that high-energy particles are beamed forward much more strongly than the standard $1/\Gamma$
characteristic from relativistic aberration of an isotropic distribution (red dashed lines). $p_0 = 
\Gamma_{\rm cm} \, \beta_{\Gamma, \rm cm}$ is the dimensionless momentum in the center-of-momentum 
frame, in which spine and layer move with equal velocity $\beta_{\Gamma, \rm cm} \, c$ in opposite 
directions. {\bf Left:} For $p_0 = 5$. {\bf Right:} For $p_0 = 15$. From \cite{Liang18}. Reproduced 
with permission from the AAS. }
\label{shearlayer_beaming}
\end{figure}

Modeling of blazar variability requires the time-dependent treatment of both the distributions
of relativistic particles and the radiation fields in the emission region. In leptonic models,
the relevant particle populations are only the electrons (and positrons, which are usually not 
distinguished from electrons, as they cool and radiate identically, and pair annihilation is
irrelevant for highly relativistic electrons); in hadronic models, in principle, electrons/positrons
need to be evolved simultaneously with ultrarelativistic protons, pions, and muons. In almost all 
models currently available in the literature, the particle momentum distributions are assumed to be 
isotropic in the rest frame of the high-energy emission region. This is a critical simplifying assumption
which makes the models tractable, as one needs to keep track only of the particles' energy distributions
(in one dimension), and current models based on isotropic particle distributions have met with significant 
success in representing SEDs and variability patterns of blazars. However, realistically, neither relativistic
shock acceleration \citep[e.g.,][]{SB12} nor particle acceleration at relativistic shear layers 
\citep{Liang18} appear to produce isotropic particle distributions in any frame. In particular,
\cite{Liang18} have shown that relativistic shear layer acceleration produces highly beamed particle 
distributions in the direction of the shear flow, possibly leading to much more strongly beamed 
radiation patterns than the standard $1/\Gamma$ scaling resulting from relativistic aberration of
emission produced isotropically in an emission region moving at Lorentz factor $\Gamma$
(see Fig. \ref{shearlayer_beaming}). 

An interesting contribution to the discussion about anisotropy of particle distributions was published
by \cite{SL19}. These authors argue that gygoresonant pitch-angle scattering, which might isotropize 
accelerated particles in the emission region, is effective only out to some isotropization energy 
$\gamma_{\rm iso}$, typically much smaller than the break energy $\gamma_b$ due to inefficient acceleration
and/or escape. Particles with $\gamma_e > \gamma_b$ are expected to move primarily along the magnetic
field and will therefore not contribute to synchrotron emission. However, particles in the range $\gamma_{\rm iso}
< \gamma_e < \gamma_b$ may still efficiently contribute to Compton scattering. This removes the need for 
strongly sub-equipartition magnetic fields, which contradict the paradigm of dynamically important B-fields
in the jets of AGN. 

Accepting the isotropic particle distribution approximation, as done in most of the current literature,
the evolution of particle energy distributions is usually done by means of an isotropic Fokker-Planck 
equation of the form (see, e.g., \cite{MK97,Kirk98,KM99,LK00,Kusunose00,BC02,DB14,Asano14} for leptonic 
models, and \citep{Dimitrakoudis12,PM12,WS15,Diltz15} for hadronic models):

\begin{equation}
{\partial n_i (\gamma, t) \over \partial t} = Q_i (\gamma, t) - {\partial \over \partial \gamma} \left(
\dot\gamma \, n_i [\gamma, t] \right) - {n_i (\gamma, t) \over t_{\rm esc}} + {\partial \over \partial\gamma}
\left( D[\gamma] \, {\partial n_i [\gamma, t] \over \partial\gamma}\right)
- {n_i (\gamma, t) \over \gamma \, t_{\rm decay}} .
\label{FP}
\end{equation}

Here, $i$ indicates the particle species (electrons/positrons, protons, pions, muons). $Q_i (\gamma, t)$ is
an injection term, which is often used to describe rapid particle acceleration, such as first-order Fermi
acceleration. This is because the first-order Fermi acceleration time scale increases with energy as 
$t_{\rm acc, 1} \propto \gamma^{\alpha}$, typically with $\alpha \ge 1$, while radiative cooling time scales 
(at least for synchrotron and Compton scattering) scale as $t_{\rm cool} \propto \gamma^{-1}$. Thus, for 
energies below the maximum energy where $t_{\rm acc, 1} (\gamma_{\rm max}) = t_{\rm cool} (\gamma_{\rm max})$,
typically $t_{\rm acc} << t_{\rm cool}$. Thus, first-order fermi acceleration is well approximated by an 
instantaneous injection term. Depending on particle species, $Q_i (\gamma, t)$ may also include pair 
production by $\gamma\gamma$ absorption and particle production through the decay of pions or muons, 
thus directly coupling to the evolution of those parent particles. The term $\dot\gamma$ describes
systematic energy losses or gains (if not already included in $Q_i$), in particular radiative losses.
$t_{\rm esc}$ is the (possibly energy-dependent) escape time scale. The fourth term on the right-hand side 
describes diffusion in momentum space, leading to second-order Fermi acceleration where $D(\gamma)$ is the 
momentum diffusion coefficient. The last term describes the decay of unstable particles (pions, muons). 

For a self-consistent solution, in the case of hadronic models, the Fokker-Planck equations (\ref{FP}) for 
all species have to be solved simultaneously, along with the radiation transfer problem (see below). The 
most efficient way to achieve stable numerical solutions to Eq. (\ref{FP}) is through implicit Crank-Nichelson
schemes \citep[e.g.,][]{CC70,Chen11}. 

There are two main approaches to solving the radiation transfer problem in most blazar emission codes.
Most commonly employed are direct solutions to a photon continuity equation of the form

\begin{equation}
{\partial n_{\rm ph} (\epsilon) \over \partial t} = {4 \pi \, j_{\epsilon} \over \epsilon \, m_e c^2}
- \alpha_{\epsilon} \, c \, n_{\rm ph} (\epsilon) - {n_{\rm ph} (\epsilon) \over t_{\rm esc}}.
\label{photon_continuity}
\end{equation}
Here, $\epsilon = h \nu / (m_e c^2)$ is the dimensionless photon energy, $j_{\epsilon}$ is the emissivity
due to the various radiation mechanisms, $\alpha_{\epsilon}$ is the absorption coefficient, primarily due
to synchrotron self-absorption and $\gamma\gamma$ absorption (wich then feeds back into the electron/positron
Fokker-Planck equation through pair production), and $t_{\rm esc}$ is the photon escape time scale. Such schemes
are typically numerically inexpensive, but they are appropriate only for very simple (typically homogeneous,
one-zone) geometries. There are, however, a few attempts to apply such schemes also to inhomogeneous multi-zone 
models, in particular shock-in-jet \citep{JB11,Joshi14} and extended-jet \citep{PC12,PC13,RS16} models. 

An alternative method to solve the radiation transfer problem is through Monte-Carlo simulations 
\citep[e.g.,][]{Chen11,Chen12,Chen15,Zhang15,Zhang16a}. Such schemes are much more flexible in terms 
of geometries, they allow straightforward time-tagging of photons and polarization-dependent ray tracing
\citep{Zhang15,Zhang16a}. However, time-dependent multi-zone simulations quickly become extremely time
consuming due to the large number of photons that need to be tracked in order to achieve meaningful 
photon statistics. 

\begin{figure}[H]
\centering
\includegraphics[width=7cm]{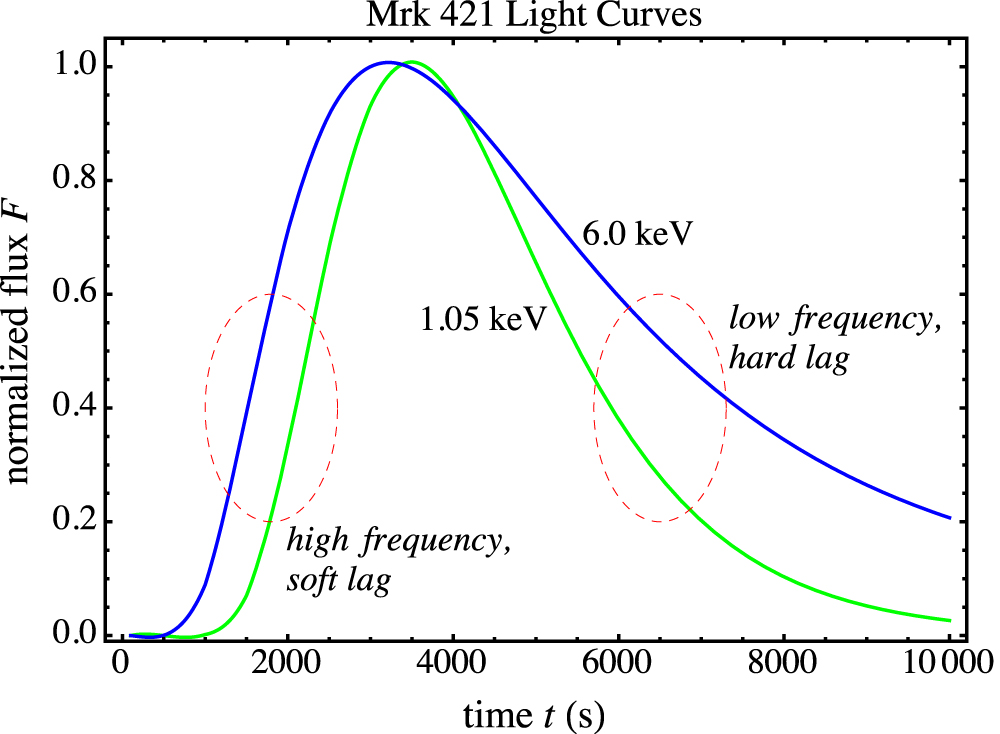}
\quad
\includegraphics[width=7cm]{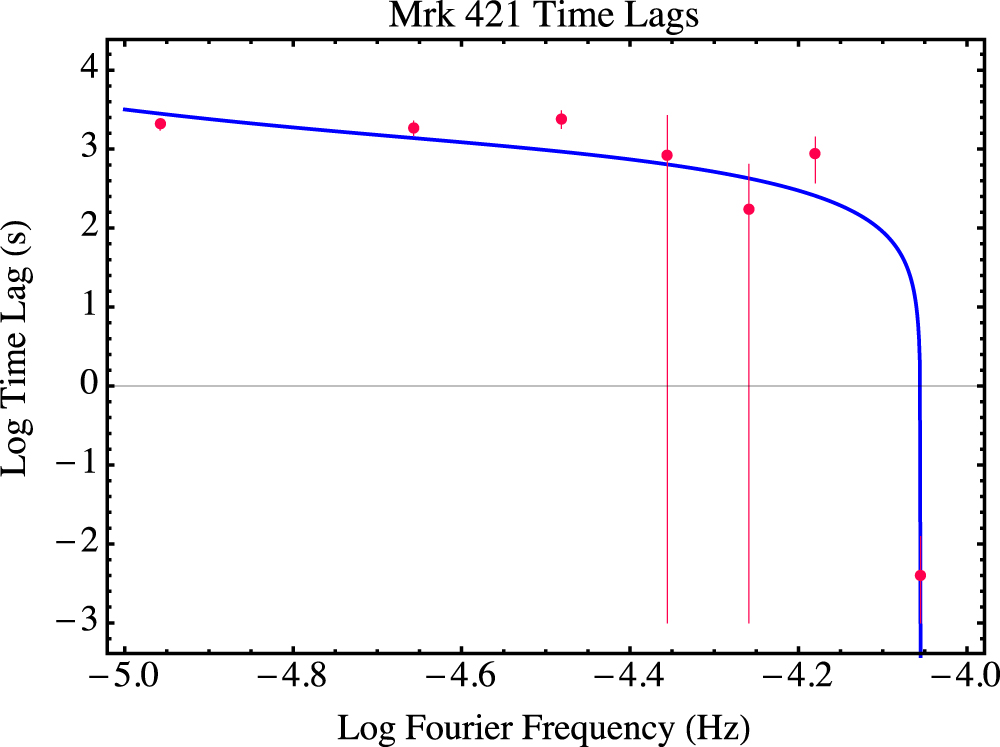}
\caption{{\it Left:} Re-constructed X-ray light curves at 1.05 and 6 keV from a Fourier-space solution to 
time-dependent electron acceleration and synchrotron emission in Mrk 421. {\it right} The resulting 
Fourier-frequency-dependent time lags, compared to {\it Beppo}SAX data from \cite{Zhang02}. 
From \cite{Lewis16}. Reproduced with permission from the AAS. }
\label{Lewis_lags}
\end{figure}   

An innovative new approach to solving time-dependent particle and photon evolution in blazar models
has been developed by \cite{FB14,FB15,Lewis16}. These authors solve the Fokker-Planck and radiation 
transfer equations in Fourier space and re-convert them into observable light curves and cross 
correlations. With this approach, it was, for the first time, feasible to model Fourier-frequency-dependent
time lags between hard and soft X-rays, as observed from Mrk 421 (see Fig. \ref{Lewis_lags}).

\subsection{\label{theo_polarization}Multi-Wavelength Polarization Modeling}

In this section, various models advanced to explain the large-angle optical PA swings in flaring blazars
as well as predictions for future high-energy polarimeters will be discussed.

\subsubsection{Optical polarization angle swings}

The large-angle optical PA rotations associated with multi-wavelength flares discussed in Section
\ref{obs_polarization} have spurred a large number of theoretical works to interpret these events.
The large degree of optical polarization is a clear indication that the non-thermal emission is
synchrotron radiation in partially ordered magnetic fields. A non-thermal synchrotron spectrum with 
energy index $\alpha = (p - 1)/2$, where $p$ is the underlying non-thermal electron spectral index,
can be maximally polarized by a degree of

\begin{equation}
\Pi_{\rm max} = {p + 1 \over p + 7/3} = {\alpha + 1 \over \alpha + 5/3}
\label{Pi_max}
\end{equation}
in the case of a perfectly ordered magnetic field. For typical spectral indices of $p \sim 2$ -- 3, this
corresponds to $\sim 70$ -- 75~\% polarization. The observed degree of optical polarization is typically
in the range $\Pi_o \sim$~a few -- 30~\%, this indicates that the magnetic fields in the optical emission regions
must be partially ordered. Changes in the PA then likely indicate a change in the orientation of the magnetic 
field with respect to our line of sight. This can be either an intrinsic change in a (more or less) straight 
jet, or it can indicate a change of the jet orientation with respect to the line of sight. Alternatively, PA 
variations may be the result of stochastic processes in a turbulent jet environment. All of these possibilities
will be discussed in more detail below. 

\begin{figure}[H]
\centering
\hbox{\includegraphics[width=7cm]{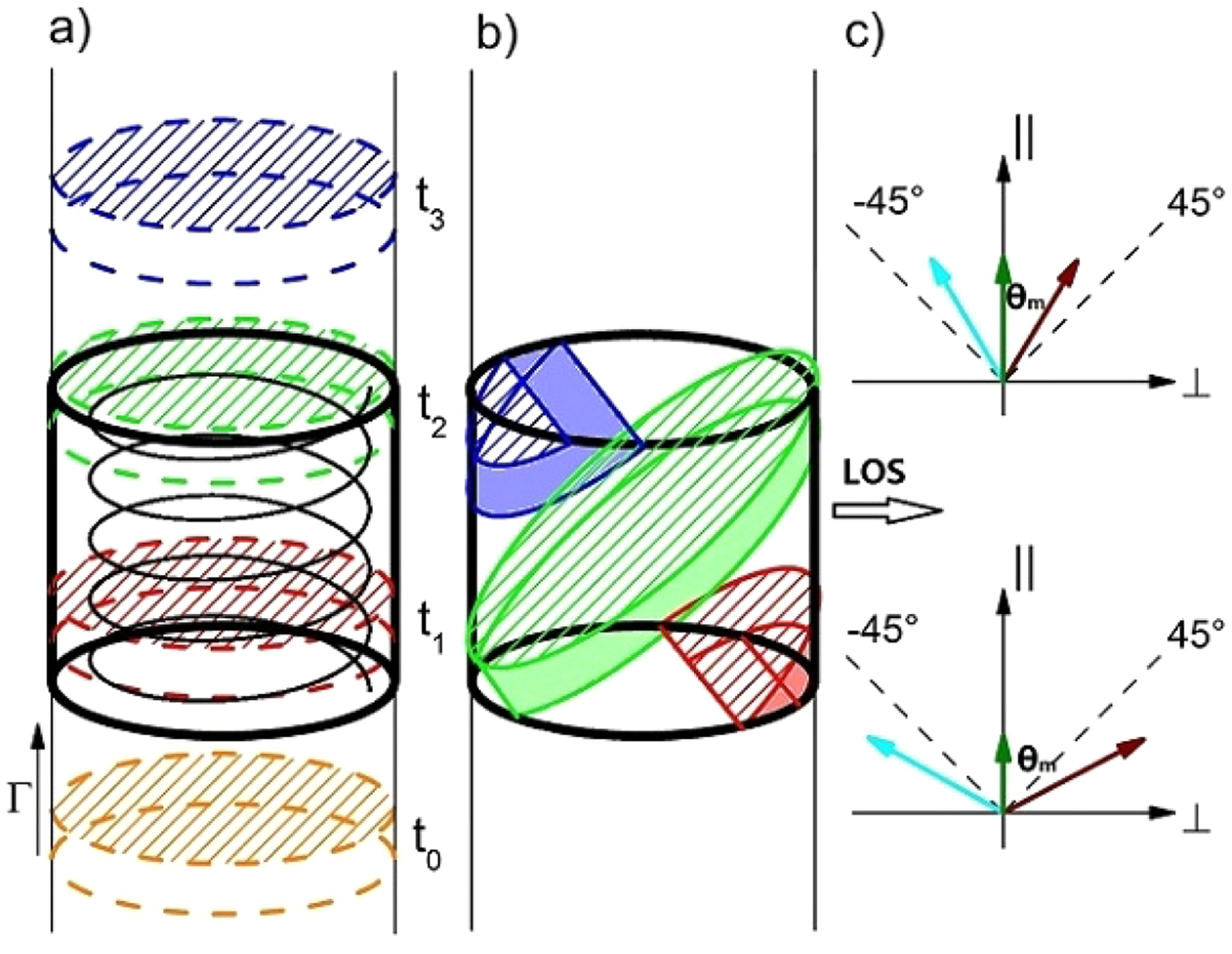}
\quad
\includegraphics[width=7cm]{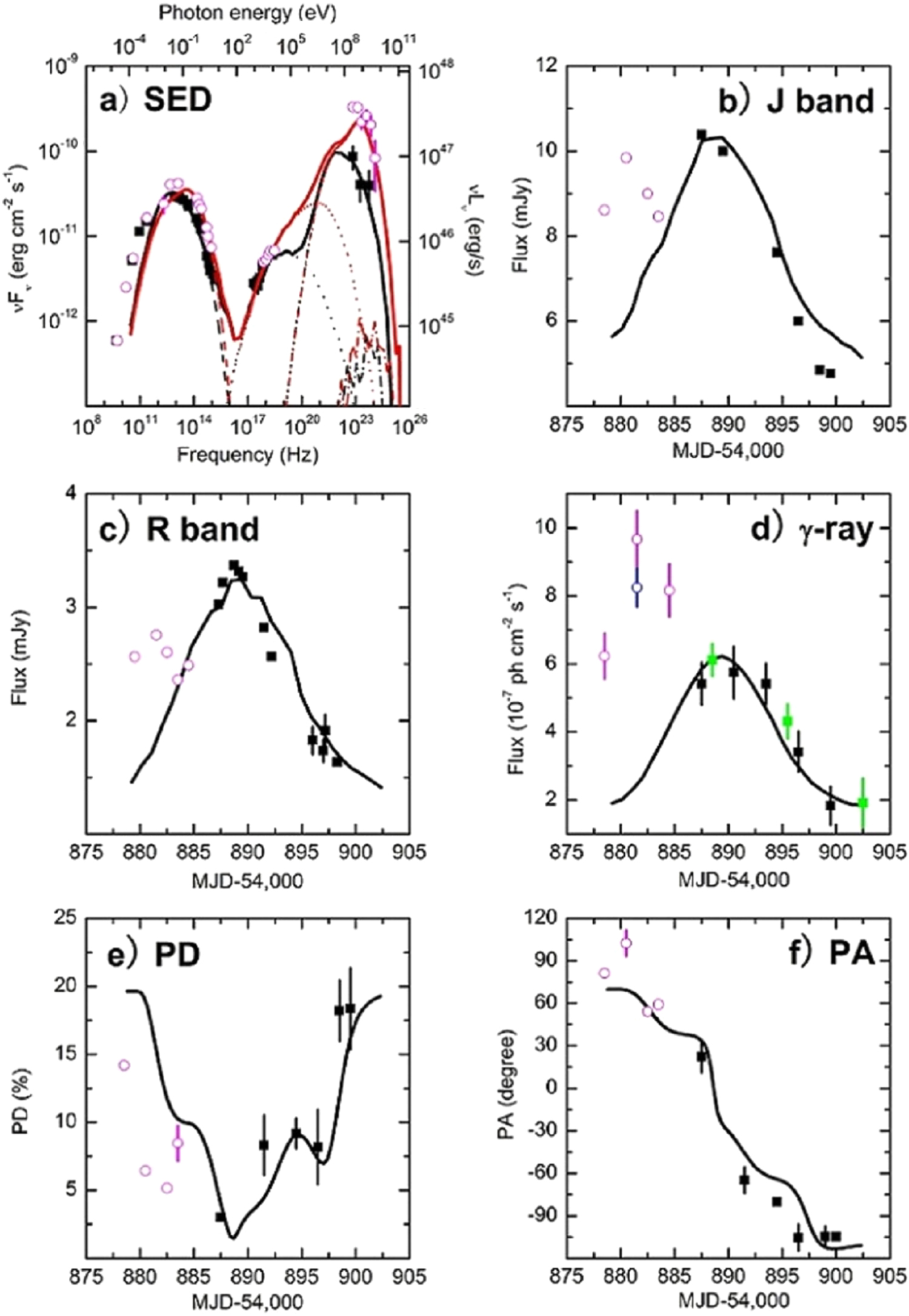}}
\caption{{\it Left:} Sketch of the magnetic-field geometry and illustration of light-travel-time effects in
a shock-in-jet model with helical magnetic field. Different colors indicate the location of the shock front 
at different times: In the left-most sketch, equal times in the AGN rest frame; in the middle sketch: Location
of the shock front at equal photon-arrival times at the observer, illustrating how different parts of a helical
magnetic fields are ``lit-up'' by the shock front, as seen by the observer at different times, leading to a
gradually changing dominant magnetic-field (and, thus, polarization) direction. 
{\it Right:} The resulting (a) snap-shot SEDs, (b) -- (d) multi-wavelength light curves, (e) polarization 
degree, and (f) polarization-angle swing compared to observations of 3C279 by \cite{Abdo10b}. 
From \cite{Zhang15}. Reproduced with permission from the AAS. }
\label{3C279swing}
\end{figure}  

Intrinsic magnetic field changes may be caused through magnetic-field compression in a shock. Models based on
shock propagation in a jet pervaded by a magnetic field have been advanced, in particular, by 
\cite{Zhang14,Zhang15,Zhang16a}. Here, the finite transverse light travel time plays a crucial role 
in producing PA swings. \cite{Zhang15} have used such a model to self-consistently explain SEDs, 
multi-wavelength light curves, and PA and $\Pi$ variations, including a $\sim 180^o$ PA rotation, 
in the FSRQ 3C279, as observed by \cite{Abdo10b} (see Fig. \ref{3C279swing}). A potential drawback 
of such a model is that a single shock passage predicts swings of at most $180^o$. Thus, rotations by multiples
of $180^o$ would require a succession of multiple shocks. Furthermore, assuming that the helicity of the magnetic
field structure does not change in the same object between different observation periods, such rotations are 
predicted to occur always in the same direction, which is contrary to observations. Thus, while this model has
been very successful in explaining some PA rotations associated with multi-wavelength flares, it can likely not 
be applied to all.

\begin{figure}[H]
\centering
\hbox{\includegraphics[width=7cm]{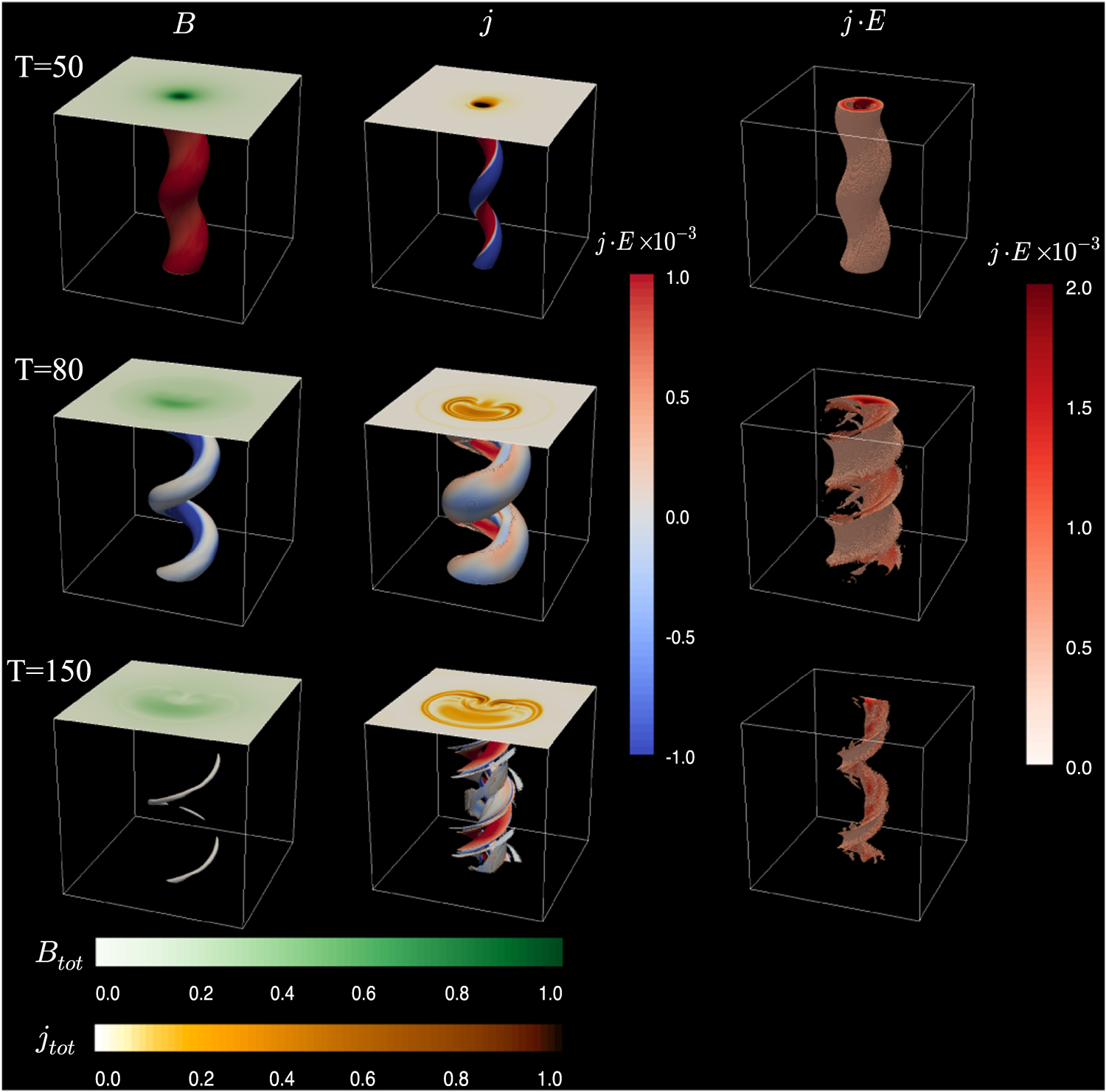}
\quad
\includegraphics[width=6cm]{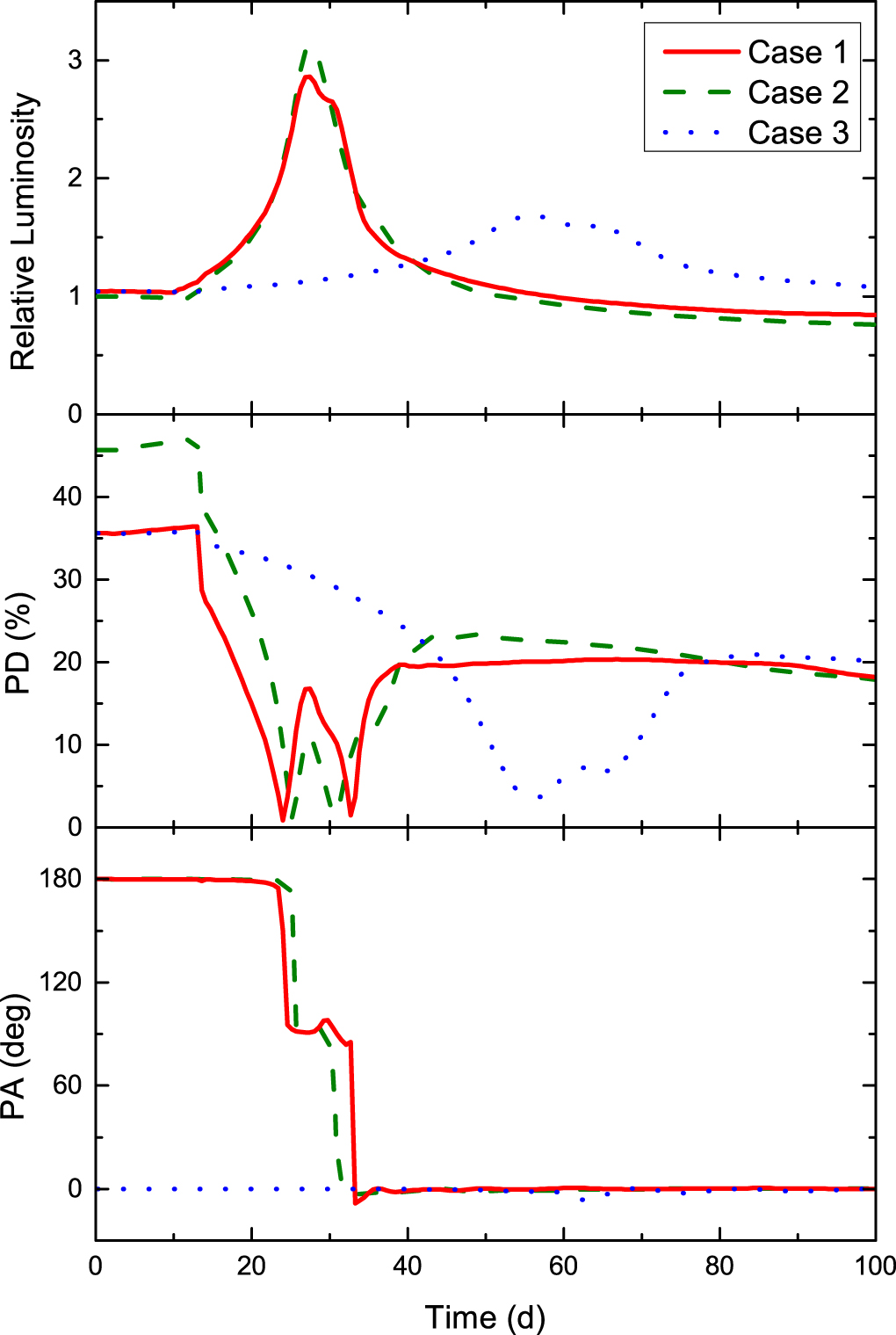}}
\caption{{\it Left:} MHD simulations of the evolution of the kink instability in a blazar jet. {\it Right:} The 
resulting (a) light curves, (b) polarization degree, and (c) polarization-angle swing. The simulation in the left 
panel corresponds to Case 1 in the right panel, corresponding to a magnetization of $\sigma_m = 2$. For Case 2, 
the initial helical B-field is radially more strongly confined; for Case 3, the magnetization is lower ($\sigma_m
= 0.2$) compared to Case 1. From \cite{Zhang17}. Reproduced with permission from the AAS. }
\label{kink_swing}
\end{figure}

Models along similar lines involve magnetic-field re-structuring and particle energization through the
kink instability \citep{Nalewajko17,Zhang17} (see Fig. \ref{kink_swing}). This process will also lead 
to flaring activity, with the strength of flares depending on the initial magnetization, correlated with
PA swings. In this model, as with the internal-shock model, the unit of PA rotations is $180^o$, also 
expected to occur always in the same direction.

\begin{figure}[H]
\centering
\includegraphics[width=7cm]{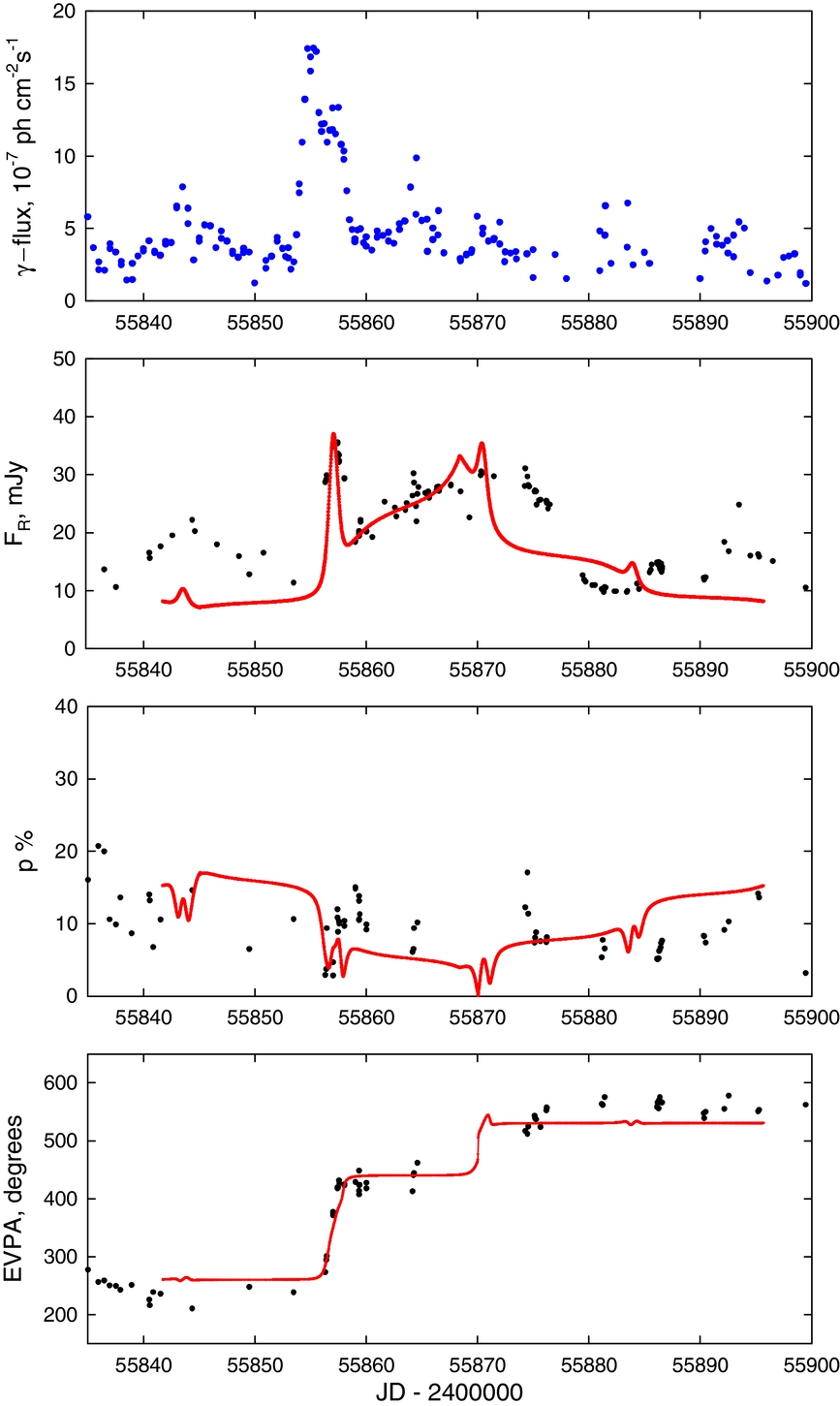}
\caption{Fits to the optical light curve, polarization degree, and PA of S5 0716+714 in October 2011,
using a model of a shock in a helical jet. 
From \cite{Larionov13}. Reproduced with permission from the AAS. }
\label{Larionov0716}
\end{figure}

Alternatively, PA swings may result from a change in the jet orientation, such as assumed in the helical-jet
model \citep[e.g.,][]{VR99,Ostorero04,Marscher08,Larionov13,Larionov16,LK17,Raiteri17}. In particular, 
\citep{Larionov13,Larionov16} modelled the optical flux and polarization variability of S5 0716+71 and 
CTA 102, respectively (see Fig. \ref{Larionov0716}). Such models have been successful in representing flux 
and polarization variability in some cases. However, also here the direction of the rotations is expected 
to be pre-determined by the helicity of the jet (and B-field structure) and thus not changing for a given 
object. Also, as discussed in Section \ref{theo_variability}, at least in their simplest form, they predict 
essentially achromatic multi-wavelength variability, which is rarely observed. 

\begin{figure}[H]
\centering
\includegraphics[width=7cm]{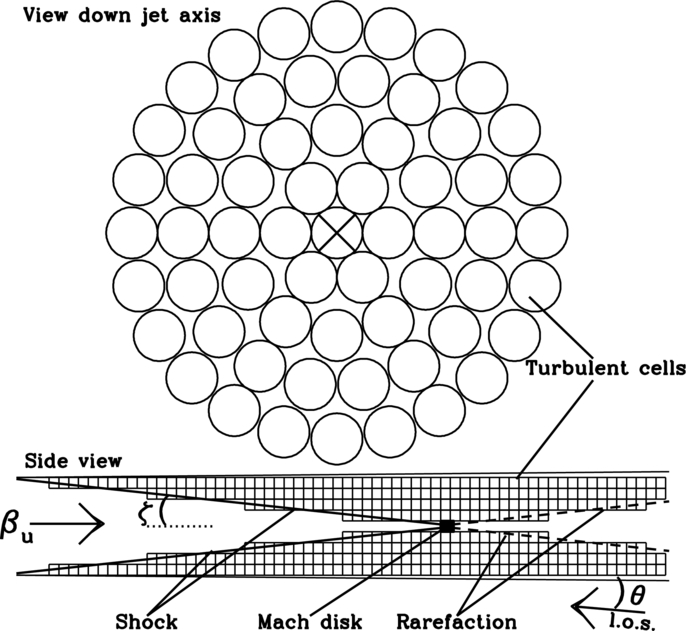}
\quad
\includegraphics[width=7cm]{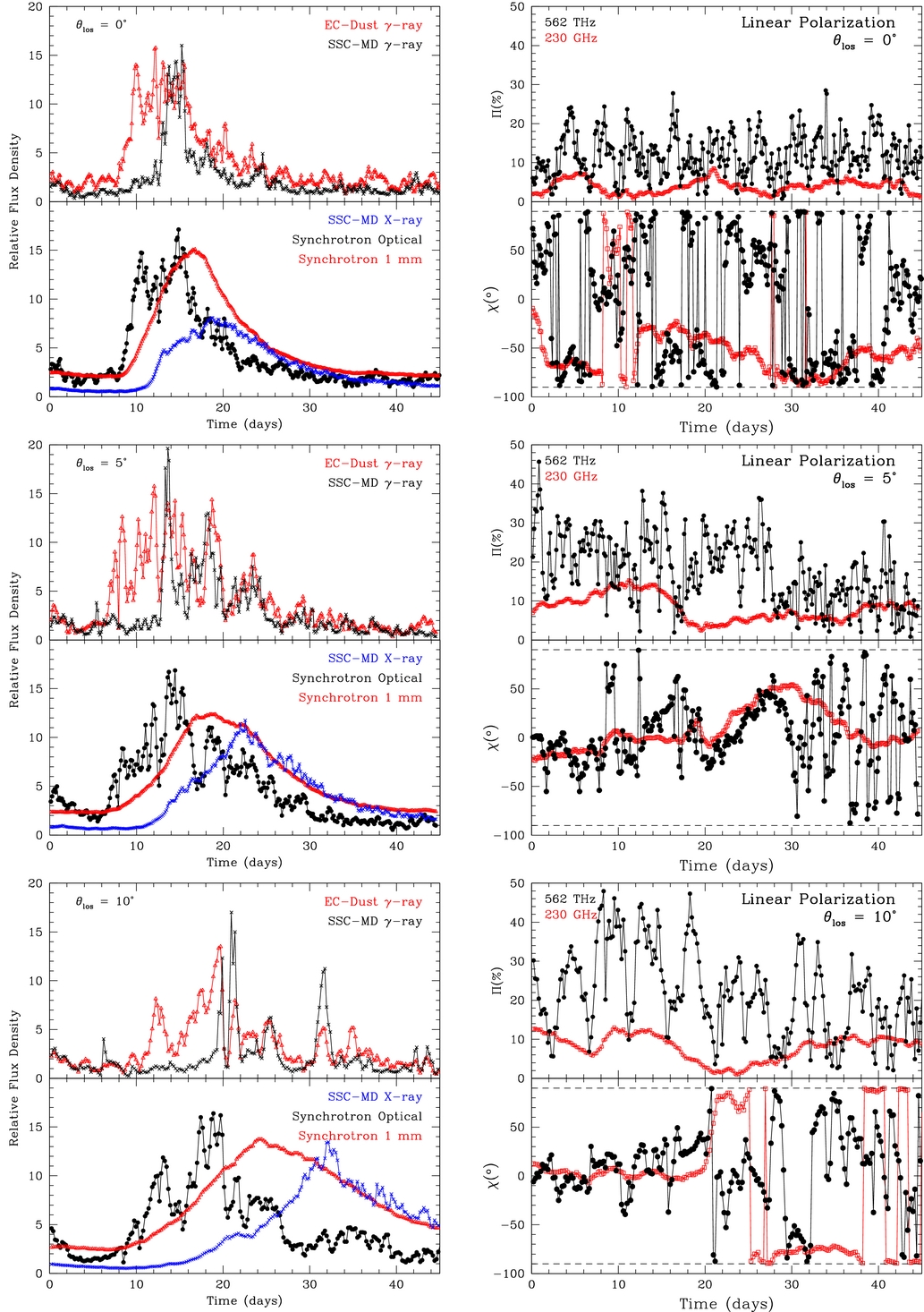}
\caption{{\it Left:} Sketch of the Turbulent Extreme Multi-Zone model. {\it Right} Simulated light curves
(left) and polarization variability (right) for a representative test case of the TEMZ model, for three 
different viewing angles. From \cite{Marscher14}. Reproduced with permission from the AAS. }
\label{TEMZ}
\end{figure}

Finally, stochastic polarization variability may result from turbulent environments in the jet. In
particular, \cite{Marscher14} developed a ``Turbulent Extreme Multi-Zone'' (TEMZ) model in which 
different turbulent cells in the jet are characterized by different magnetic-field orientations.
The summed radiation of a large number of such turbulent cells will result in a (normally small)
residual polarization with stochastically varying direction. This model has successfully reproduced
stochastic polarization and flux variations in blazars (see Fig. \ref{TEMZ}) and may occasionally 
also lead to large-angle PA swings. It naturally accounts for changing directions of PA swings 
in the same object and has also been used to model 
circular radio polarization from the inner jet regions of blazars \citep{MM18}. However, due to
the stochasticity of both flux and polarization variations, large-angle PA swings are expected to
be very rare, and such swings are not expected to correlate systematically with multi-wavelength
flares, as observed by the RoboPol experiment \cite{Blinov18}.

\subsubsection{High-energy polarization}

As will be discussed in more detail below in Section \ref{prospects}, there are currently great
prospects for future detections of high-energy (X-ray and $\gamma$-ray) polarization from blazars.
Thus, it is timely to consider model predictions of such high-energy polarization. In addition to
synchrotron radiation, also Compton scattering may induce polarization, but only in the case of
scattering off non-relativistic electrons \citep[see, e.g.,][]{Krawczynski12}. Inverse-Compton
scattering by relativistic electrons does not induce polarization, but reduces the degree of 
polarization of a polarized target photon field by at least $\sim 1/2$ \citep[e.g.,][]{BS73}.

\begin{figure}[H]
\centering
\includegraphics[width=12cm]{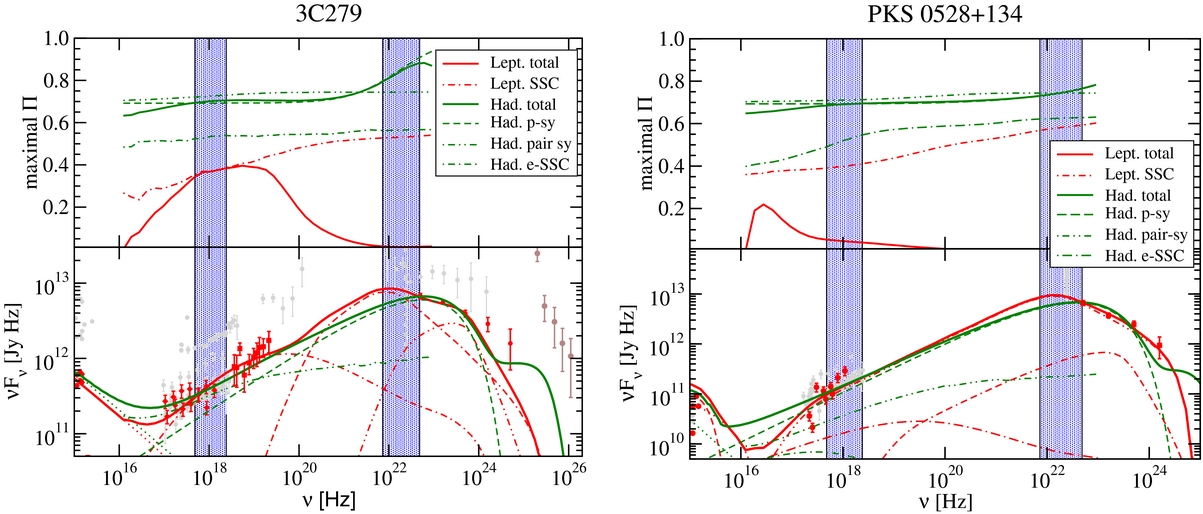}
\caption{Predictions of X-ray and $\gamma$-ray polarization for two FSRQs. {\it Bottom:} Zoom-in
on the UV -- $\gamma$-ray SED, based on data from \citep{Abdo10} and leptonic (red) and hadronic
(green) SED fits from \cite{Boettcher13}. {\it Top:} Predicted high-energy polarization from the
SED models of the bottom panel. The vertical shaded bands indicate the 2 -- 10~keV X-ray and and
the 30 -- 200~MeV $\gamma$-ray band. From \cite{ZB13}. Reproduced with permission from the AAS. }
\label{Xgpolarization}
\end{figure}

Detailed predictions for the X-ray and $\gamma$-ray polarization in blazars have been made by
\cite{ZB13} (see Fig. \ref{Xgpolarization}). In the case of leptonic models, the X-ray emission 
in blazars is generally dominated by either synchrotron (in HSPs) or SSC 
emission (in ISPs and LSPs) and thus expected to be polarized. The $\gamma$-ray emission in LSP (and
ISP) blazars is usually dominated by Compton scattering of external radiation fields by relativistic 
electrons and, thus, unpolarized, whereas in HSP blazars, it is usually modelled as being due to 
SSC emission, thus exhibiting a low, but non-zero, degree of polarization. In hadronic models, 
the high-energy emission is dominated by synchrotron emission of either protons or secondary
pairs from photo-pion production and subsequent cascades, and thus expected to be polarized 
with a similar degree of polarization as the optical. High-energy polarization can thus be
used as a diagnostic between leptonic and hadronic models.

\begin{figure}[H]
\centering
\includegraphics[width=12cm]{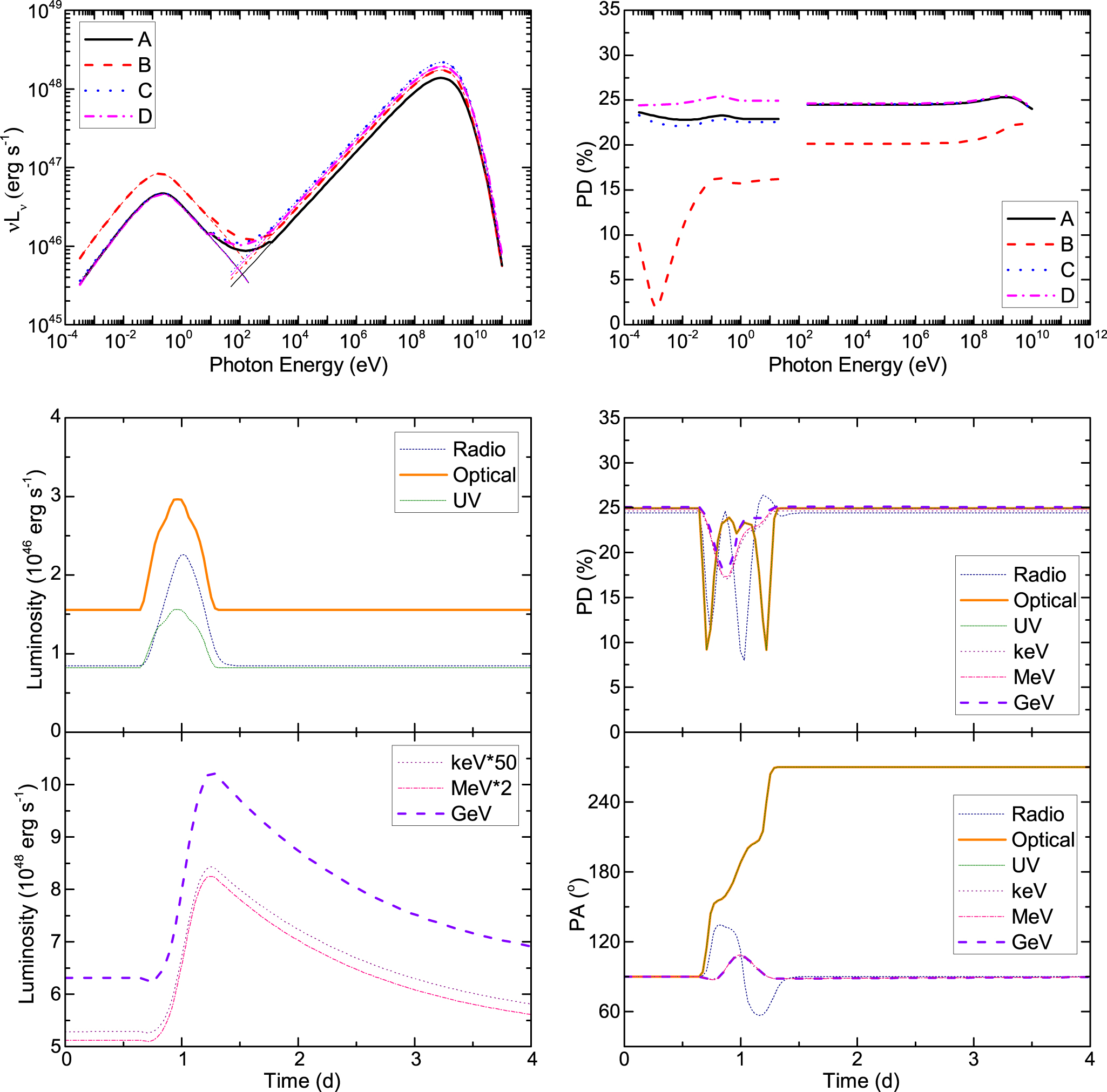}
\caption{Representative simulations of a hadronic shock emission model with polarization predictions.
{\it Top left:} Snap-shot SEDs, {\it Middle and bottom left:} Multi-wavelength light curves, 
{\it Top right:} Frequency-dependent polarization degree at different times, {\it Middle right:}
Time-dependence of polarization degree at various frequencies, {\it Bottom right:} Time-dependent 
PA at various frequencies. The figure illustrates that in a hadronic shock emission model, optical 
(primary electron synchrotron) polarization-angle swings are not expected to be accompanied by 
similar PA swings at X-rays and $\gamma$-rays. From \cite{Zhang16b}. Reproduced with permission from the AAS. }
\label{Hadron_PAswing}
\end{figure}  

A first attempt at a time-dependent, multi-zone hadronic model with polarization-dependent radiation
transfer has been published by \cite{Zhang16b}. The model is applicable in a parameter regime in 
which the high-energy emission is dominated by proton synchrotron radiation, as appears to be preferred, 
at least for most LSPs when fitted with hadronic models \citep[e.g.,][]{Boettcher13,Petropoulou15}.
\cite{Zhang16b} show that, in such a hadronic scenario, even though optical PA swings may be produced 
by a shock in a jet with a helical magnetic field, similar PA swings are not expected in the X-ray 
and $\gamma$-ray polarization (see Fig. \ref{Hadron_PAswing}). 
This is because of the much longer radiative cooling time of protons
responsible for the X-ray and $\gamma$-ray emission, which therefore occupy a much larger active volume
than the optical-synchrotron emitting primary electrons, so that the effect of the shock on the 
magnetic field orientation is less pronounced for the proton synchrotron emission. Consequently, the 
X-ray and $\gamma$-ray PA is expected to remain relatively stable even during a shock-induced flare,
which significantly improves the chances of experimental detection of such polarization.

\subsection{\label{theo_multimessenger}Models of Neutrino Emission from Blazars}

Blazars have long been considered a prime candidate for the source of at least part of the VHE neutrinos 
detected by IceCube 
\citep[e.g., 
][]{MB89,Stecker91,PS92,MB92,Mastichiadis96,Muecke03,Reimer05,Dimitrakoudis12,Halzen13,Murase14,Dermer14,Petropoulou15b}.
In hadronic models of blazars, protons are accelerated to sufficiently high energies to produce $\gamma$-ray
emission via proton synchrotron radiation and/or photo-pion production, $p + \gamma \to p + \pi^o$, 
$p + \gamma \to n + \pi^+$, or higher-order processes with multi-pion production, followed by pion decay,
$\pi^{\pm} \to \mu^{\pm} + \nu_{\mu} (\overline{\nu_{\mu}})$ and muon decay, $\mu^{\pm} \to e^{\pm} + 
\overline{\nu_{\mu}} (\nu_{\mu})  + \nu_e (\overline{\nu_e})$. In typical AGN jet environments, photo-pion
production is expected to be significantly more efficient than hadro-nuclear (i.e., proton-proton)
interactions. Therefore, almost all works on neutrino production in AGN work on the basis of photo-pion 
production \citep[see, however, hadro-nuclear emission models for neutrino production in AGN 
by ][]{Liu18a,Liu18b}. Usually, $p\gamma$ interactions are 
strongly dominated by single-pion production through the $\Delta^+$ resonance at an energy of
$E_{\Delta^+} = 1232$~MeV, in which case each photo-pion interaction results in the production 
of 3 neutrinos, each carrying an average energy of $\sim 5$~\% of the proton energy, 

\begin{equation}
E'_{\nu} \approx 0.05 \, E'_p
\label{Enu}
\end{equation}
where the prime indicates quantities in the rest frame of the emission region, in which the interactions
are assumed to take place. Interaction at the $\Delta^+$ resonance energy requires that the photon  
($E'_{\gamma}$) and proton energies obey the relation for the center-of-momentum energy squared, $s$,

\begin{equation}
s \approx E'_{\gamma} \, E'_p \approx E_{\Delta^+}^2
\label{scm}
\end{equation}

Hence, combining Equations (\ref{Enu}) and (\ref{scm}), one finds that the production of IceCube neutrinos 
of energies $E_{\nu} = \delta \, E'_{\nu} = 100 \, E_{14}$~TeV requires protons of energy 

\begin{equation}
E'_p \approx 200 \, E_{14} / \delta_1 \, {\rm TeV}
\label{Ep}
\end{equation}
and target photons of energy 

\begin{equation}
E'_{\gamma} \gtrsim 1.6 \, \delta_1 \, / E_{14} \, {\rm keV}.
\label{Etarget}
\end{equation}

Hence, first, while the sources of IceCube neutrinos must be able to accelerate protons to $\sim$~EeV energies,
they are not necessarily the sources of ultra-high-energy cosmic rays (UHECRs, with energies $E_{\rm UHECR}
> 10^{19}$~eV). Second, Equation (\ref{Etarget}) indicates that, for efficient IceCube neutrino production,
an intense target photon field at X-ray energies (in the co-moving frame of the emission region) is required. 
Note that the co-moving primary-electron-synchrotron radiation field in all blazars (especially, LSP and ISP 
blazars) peaks at much lower energies than required by Equ. (\ref{Etarget}). Therefore, photo-hadronic blazar 
models utilizing the co-moving synchrotron radiation fields typically produce very hard IceCube neutrino spectra,
peaking significantly above the IceCube energy range \citep[see, e.g.,][]{Cerruti18}. 

Further constraints on the photo-hadronic scenario to produce both high-energy $\gamma$-rays and PeV neutrinos
stem from the fact that the target photon field for photo-pion production also acts to absorb any co-spatially 
produced $\gamma$-rays via $\gamma\gamma$ absorption. The $p\gamma$ cross section is several orders of magnitude
smaller than the $\gamma\gamma$ absorption cross section. Efficient $p\gamma$ neutrino (and $\gamma$-ray)
production requires that the optical depth for relativistic protons to interact with the target photon field, 
$\tau_{p\gamma} \sim 1$, which then implies that the optical depth of the emission region to $\sim$~GeV photons
is $\tau_{\gamma\gamma} \sim 310 \, \tau_{p\gamma} \gg 1$ \citep[e.g.,][]{Reimer18}. Thus, any $\gamma$-rays 
produced co-spatially with IceCube neutrinos, are expected to be strongly absorbed and initiate electromagnetic 
cascades, whose energy will ultimately escape the emission region in the optical -- UV -- X-ray regime. 

This led several authors \citep[e.g.,][]{Gao18,Murase18,Keivani18}, to conclude that, if blazars are the 
sources of (at least some) IceCube neutrinos, their $\gamma$-ray emission is likely to be dominated by 
leptonic processes, and no correlation between $\gamma$-ray and neutrino emissions is necessarily expected. 
Instead, the unavoidable cascade emission from photo-hadronic processes in blazar jets is expected to leave 
an imprint in the X-ray emission from blazars, which may be a better indicator of neutrino-production
activity than $\gamma$-rays.

\begin{figure}[H]
\centering
\includegraphics[width=12cm]{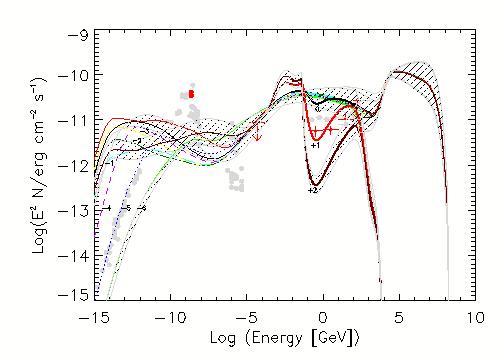}
\caption{Simulations of synchrotron-supported cascades with target photon and proton spectra appropriate 
to produce the neutrino flux from TXS 0506+056 during the 2014 -- 2015 neutrino flare, compared to contemporaneous
optical, X-ray (upper limit from {\it Swift}-BAT) and GeV $\gamma$-ray ({\it Fermi}-LAT) data. The different 
curves (different colors) are labelled by the log of the maximum $\gamma\gamma$ optical depth, 
$\log(\tau_{\gamma\gamma}^{\rm max}$, and the shaded areas indicate error margins based on the 
error on the measured IceCube neutrino spectrum. All cases violate either the X-ray or $\gamma$-ray 
constraints. From \cite{Reimer18}.}
\label{cascades}
\end{figure}  

This conclusion is further corroborated by detailed studies of electromagnetic cascades initiated by 
photo-hadronic neutrino-production processes in blazar jet environments by \cite{Reimer18}. \cite{Reimer18}
investigated possible regimes of electromagnetic cascades based on the energetics requirements to produce
the neutrino flux from TXS 0506+056 during the 2014 -- 2015 neutrino flare, and found that a synchrotron-supported
cascade regime can be ruled out, as the cascades violate observational constraints (see Fig. \ref{cascades}). 
Further considering the nature of the target photon field, they found that there is only a narrow range of 
parameters not violating observational constraints, requiring a $p\gamma$ UV -- soft X-ray target photon
field external to the jet (stationary in the AGN rest frame) and a kinetic luminosity in protons near the 
Eddington limit of the central black hole in TXS 0506+056. The nature of the required UV -- soft X-ray target 
photon field remains unclear, but could possibly be related to a radially structured jet
\citep[spine-sheath, see][]{GTC05,TG08,Ansoldi18}, or a radiatively inefficient accretion flow \citep{Righi18}.

To conclude, photo-hadronic PeV neutrino production in blazar jets is not necessarily correlated with 
$\gamma$-ray activity (but rather X-ray activity), and due to the expected high $\gamma\gamma$ opacity 
of the required $p\gamma$ target photon fields, $\gamma$-rays in neutrino-producing blazars are likely 
to be dominated by leptonic processes, likely not spatially coincident with the site of neutrino production.

\section{\label{prospects}Future Prospects}

The next decade will witness the start of operations of 
several new ground- and space-based astronomy facilities.
Of particular relevance to blazar research will be, in the
author's opinion, the Cherenkov Telescope Array
\citep[CTA][]{Acharya13} and the Imaging X-ray Polarimetry
Explorer \citep[IXPE, ][]{Weisskopf16}, while ground-based
radio and optical flux and polarimetry monitoring projects
will continue, and hopefully the {\it Fermi} Gamma-Ray
Space Telescope and the Neil Gehrels {\it Swift} X-Ray
Observatory will continue to provide all-sky GeV
$\gamma$-ray monitoring and flexible X-ray coverage to
blazar observations, respectively, for several years to
come. KM3NeT \citep{Margiotta14,Adrian16b} and planned upgrades
to IceCube \citep[IceCube-Gen2][]{Blaufuss15,Aartsen17b} and the 
Lake Baikal neutrino detector \citep{Avrorin11} will greatly improve 
our view of the neutrino sky and hopefully provide a definitive
answer whether blazars are PeV neutrino sources. If selected, 
also future missions such as the All-sky Medium Energy Gamma-ray 
Observatory \citep[AMEGO, ][]{Moiseev17} would provide a boost 
to the study of blazars.

\subsection{\label{CTA}The Cherenkov Telescope Array}

CTA\footnote{\tt http://www.cta-observatory.org/} will be the 
next-generation ground-based Cherenkov Telescope facility, 
consisting of $\sim 100$~Cherenkov telescopes of 3 different 
sizes at two sites, one in the nothern hemisphere (La Palma,
Canary Islands, Spain), and one in the southern hemisphere
(Paranal/ESO, Chile). It will improve on the sensitivity of
current IACT facilities by about an order of magnitude and
extend the observable energy range to $\sim 20$~GeV -- 
$\sim 300$~TeV. The current schedule foresees CTA to be
operational by 2024. In addition to key science projects
conducted by the CTA Consortium, CTA observing time will be
open to community through a competitive proposal process. 
For an in-depth discussion of the science potential and key 
science projects to be conducted with the CTA, see 
\cite{CTA17}. Highlights of progress that CTA promises for
blazar studies, include:

\begin{enumerate}

\item{{\bf Short-term variability:} The $\sim 10$-fold improved sensitivity of CTA 
compared to currently operating IACT facilities will allow for the study of blazar 
variability on short time-scales. In the case of the brightest flares, variability 
down to sub-minute time-scales could, in principle, be detected, if present. Already 
the $\sim 5$~min. variability observed in a few cases is severely challenging current 
blazar models, as discussed in Section \ref{theo_variability}. If blazar do indeed show 
significant variability on sub-minute time scales, this would call for a profound paradigm
shift concerning our understanding of $\gamma$-ray emission from blazars. 

The detection of minute-scale variability strongly suggests a location of the emission 
region close to the central black hole, at a distance of the order $r \sim 2 \, \Gamma^2
\, c \, t_{\rm var} \sim 3.6 \times 10^{14} \, \Gamma_1^2 \, t_{\rm var, min}$~cm. Even for bulk
Lorentz factors of order $\Gamma \sim 100$, this would normally be within the Broad Line Region
(BLR) of an FSRQ. Thus, in the case of FSRQs (such as PKS 1222+216, from which sub-hour variability 
has been seen), it would be very difficult to avoid $\gamma\gamma$ absorption by the BLR radiation 
field \citep[e.g.,][]{Liu06,PS10,Finke16,BE16}. Thus, such detections either require exotic physics
to suppress $\gamma\gamma$ absorption within the BLR \citep[e.g.,][]{Tavecchio12,TB16}, or a mechanism
to produce very compact emission regions along the jet at $\sim pc$ distances from the central black
hole. 

The improved sensitivity will also enable the detection of short-term variability in fainter 
$\gamma$-ray-detected AGN, including mis-aligned blazars (seen as radio galaxies) and narrow-line 
Seyfert-1 galaxies, some of which also appear to show blazar-like properties \citep[e.g.,][]{Abdo09,Dammando16}. 
Thus, we will be able to study whether sub-hour variability is a property of only a few blazars, or whether 
it is a common phenomenon among $\gamma$-ray emitting radio-loud AGN. 
}

\item{{\bf Detailed $\gamma$-ray spectral studies:} The improved sensitivity and extended energy 
range of CTA compared to current IACTs will enable detailed $\gamma$-ray spectral studies with 
substantial overlap in energy with the {\it Fermi}-LAT. Potential spectral features due to $\gamma\gamma$ 
absorption in the BLR or dust torus radiation fields of the AGN are expected to arise in the $\sim 10$
-- 100~GeV regime \citep[e.g.,][]{PS10}, i.e., just in the transition region between the energy ranges
of {\it Fermi}-LAT and IACTs. Currently, the study of spectral features in this regime, beyond 
simple exponential cut-offs or similar smooth spectral shapes \citep[see, e.g.,][]{Senturk13}, 
is complicated by the fact that a high-quality {\it Fermi}-LAT spectrum, for most blazars, typically 
requires exposures of (at least) several days, during which the source is likely to show substantial 
variability. On the other hand, IACTs perform short, pointed observations of at most a few hours per 
night. Thus, any spectral features in this overlap energy range may well be an artifact of the often 
vastly different integration times in the GeV and TeV regimes. The substantial energy overlap between
{\it Fermi}-LAT (up to $\lesssim 100$~GeV) and CTA ($\gtrsim 20$~GeV) will enable to proper flux 
cross-calibration of the two instruments and allow for a detailed study of spectral features in the 
overlap region, at least in cases where observations do not indicate significant variability over the 
course of the joint observations. The identification of BLR $\gamma\gamma$ absorption features would 
provide a key diagnostic for the location of the $\gamma$-ray emission region and, thus, the region 
where relativistic particles are accelerated to $> TeV$ energies. The location of the particle acceleration 
region would provide strong clues towards the nature of the acceleration mechanism. 

It has also been suggested \citep[e.g.,][]{Zech17} that hadronic emission processes might lead to
detectable spectral features in the multi-TeV spectra of blazars due to separate spectral components
from muon and pion synchrotron emission and photo-pion induced pair cascades. Such features may be 
detectable, at least for nearby blazars, with the CTA. If they are systematically identified in a
sample of blazars, they might provide a ``smoking-gun'' signatures of hadronic processes and point 
to blazars as sources of high-energy cosmic rays. 
}

\end{enumerate}

\subsection{\label{IXPE}High-Energy Polarimetry}

The Imaging X-Ray Polarimetry Explorer (IXPE)\footnote{\tt https://ixpe.msfc.nasa.gov} has been 
selected by NASA for launch in (or after) 2020 as the first dedicated high-energy polarimetry 
mission. It will provide X-ray polarimetry in the 2 -- 8~keV X-ray regime and has the capacity
to detect X-ray polarization from bright blazars within a few hours of observations. In the 
$\gamma$-ray regime, the proposed AMEGO\footnote{\tt https://asd.gsfc.nasa.gov/amego/} mission 
promises to perform, for the first time, polarimetry in the MeV regime. 

X-ray polarimetry of HSP blazars, where the X-ray emission is electron-synchrotron dominated, 
would probe the degree of ordering and dominant direction of magnetic fields in the high-energy
emission region. When compared to simultaneous optical polarimetry, this will allow for test of
the co-spatiality of X-ray and optical emission in HSPs and for a comparison of the respective
emission region sizes. As the optical emission in HSPs originates from lower-energy electrons 
than the X-ray emission, one would expect the X-ray emission region to be more confined near 
the particle acceleration site, likely embedded in a more highly ordered magnetic field than 
in the, presumably much larger, optical emission site. A significantly higher degree of
polarization in X-rays compared to optical would then be expected. Such results will allow
us to study the spatial dependence of turbulence in the jet and discriminate between different
particle acceleration mechanisms, such as diffusive shock acceleration and magnetic reconnection. 

In ISP/LSP blazars, the X-rays are likely produced by synchrotron self-Compton (SSC) emission
in leptonic emission scenarios, or by proton synchrotron and cascade synchrotron emission in
hadronic scenarios. Thus, if X-ray polarimetry reveals a degree of polarization of the order
of the optical polarization in ISPs/LSPs, this would be a strong indication of hadronic emission
\citep{ZB13}. 

Along the same lines, as disussed in Section \ref{theo_polarization}, if $\gamma$-ray polarimetry
by AMEGO shows significant (of the order of the optical degree of polarization) 
MeV polarization, hadronic emission scenarios will be strongly favoured, thus identifying blazars
as (at least) PeV proton accelerators and likely sources of IceCube neutrinos. 

As the optical polarization of blazars is known to be variable, at least on daily time scales,
both in degree and angle of polarization, one might expect that the same holds true for X-ray and
$\gamma$-ray polarization. Given that, for the X-ray fainter LSP sources, IXPE might require integration 
times of days to weeks, PA changes during the exposure might destroy any intrinsic polarization signal,
unless data analysis techniques can be developed to account for a changing PA during the exposure,
such as using the PA evolution observed in the optical, as a template for the IXPE analysis. 
However, as shown by \cite{Zhang16b}, if PA rotations are produced by a straight shock-in-jet model
with a helical magnetic field, the anticipated high degree of polarization in a hadronic model 
might be measurable without the expectation of a significant PA change.

\subsection{\label{KM3NeT}Future Neutrino Detectors}

KM3NeT\footnote{\tt http://www.km3net.org} is the next-generation neutrino telescope, to be built 
at three sites at the bottom of the Mediterranean Sea. It will improve on the sensitivity of IceCube 
by more than an order of magnitude and might therefore provide the statistics to identify individual
sources of neutrinos confidently. In addition, major upgrades are planned to the existing IceCube
detector at the South Pole (IceCube-Gen2)\footnote{\tt https://icecube.wisc.edu/science/beyond} 
and to the neutrino detector in Lake Baikal\footnote{\tt http://www.inr.ru/eng/ebgnt.html}. 

If blazars are identified systematically as a source of neutrinos by future neutrino observatories, 
it would prove their nature as cosmic-ray proton accelerators (up to at least PeV energies).
However, as discussed in Section \ref{theo_multimessenger}, a correlation with GeV -- TeV $\gamma$-ray 
emission is not expected if the neutrinos are produced via photo-pion production. As there is no 
expected constrain on the $\gamma\gamma$ opacity in the case of hadro-nuclear neutrino production,
a correlation between neutrino emission and $\gamma$-ray activity may therefore hint towards this
neutrino production channel. Joint KM3NeT neutrino detections and $\gamma$-ray monitoring with 
{\it Fermi}-LAT and CTA will allow us to establish or refute such a correlation and thus constrain
the mechanism of neutrino production.

\section{\label{summary}Summary and Conclusions}

The past few years have seen many observational discoveries on blazars, among which this article highlights
a few, such as the rapid variability at GeV and TeV energies, down to just a few minutes; large-angle optical
polarization-angle rotations, found to be systematically correlated with $\gamma$-ray and multi-wavelength 
flares, and a strong hint of the blazar TXS~0506+056 as a source of IceCube neutrinos. 

Minute-scale blazar variability severely challenges existing models for blazar emission, possibly indicating 
the prominence of small-scale structures due to magnetic reconnection. Polarization-angle swings seem to hint 
towards the presence of helical magnetic field structures, possibly, but not necessarily, associated with a
changing viewing angle. The possible TXS~0506+056 + IceCube neutrino association provides a hint towards 
hadronic particle acceleration in blazars. However, a careful study of photo-pion neutrino production suggests 
that a direct correlation between $\gamma$-ray activity and neutrino emission is not necessarily expected, 
and the cascade emission going in tandem with $p\gamma$ neutrino production is likely to show up more prominently
in X-rays than in $\gamma$-rays. 

Future observations by the CTA will allow for a more sensitive exploration of minute-scale VHE $\gamma$-ray
variability in blazars, potentially invalidating the current blazar emission paradigm. IXPE and possibly 
AMEGO promise the first X-ray (and $\gamma$-ray) polarimetry results, which will aid in
testing the co-spatiality of optical and X-ray emissions and provide further diagnostics for hadronic vs.
leptonic emission scenarios. When analysing future IXPE observations of blazars, special care has to be taken 
to account for possible PA rotations of the X-ray polarization, which might mirror those seen in the optical. 

On the multi-messenger side, the future KM3NeT will greatly improve the statistics of astrophysical PeV 
neutrinos and promises the clear identification of a source class (or multiple source classes), possibly 
including blazars. This would provide conclusive proof of PeV proton acceleration in AGN jets, but does not
necessarily point towards the origin of ultra-high-energy cosmic-rays in AGN. 

\vspace{6pt}

\authorcontributions{MB is the sole author of this publication.}

\acknowledgments{The work of M.B. is supported by the Department of Science and Technology and National Research 
Foundation\footnote{Any opinion, finding and conclusion or recommendation expressed in this material is that of 
the author and the NRF does not accept any liability in this regard.} of South Africa through the South African 
Research Chairs Initiative (SARChI), grant no. 64789.}

\conflictsofinterest{The author declares no conflict of interest.}

\reftitle{References}

\end{document}